\documentclass[showpacs,amsmath,amssymb,twocolumn]{revtex4}
\usepackage{amsfonts}
\usepackage{amssymb}
\usepackage{bm}
\usepackage{longtable}
    \makeatletter
    
    \newcommand{\Rmnum}[1]{\expandafter\@slowromancap\romannumeral #1@}
    \makeatother
\allowdisplaybreaks[4]

\begin{document}
\title{ \textbf{QCD inspired relativistic bound state model} and meson structures}

\author{Shun-Jin Wang}
\email{sjwang@home.swjtu.edu.cn} \affiliation{School of Physics and
Technology, Sichuan University, Chengdu, 610064, PR China}
\author{ Jun Tao}
\email{taojscu@gmail.com} \affiliation{School of Physics and
Technology, Sichuan University, Chengdu, 610064, PR China}
\author{Xiao-Bo Guo}
\affiliation{School of Science, Southwest University of Science and
Technology, Mianyang 621010, PR China}
\author{ Lei Li }
\affiliation{School of Science, Southwest University of Science and
Technology, Mianyang 621010, PR China}


\begin{abstract}
A QCD inspired relativistic effective Hamiltonian model for the
bound states of mesons has been constructed, which integrates the
advantages of several QCD effective Hamiltonian models. Based on
light-front QCD effective Hamiltonian model, the squared invariant
mass operator of meson is used as the effective Hamiltonian. The
model has been improved significantly in four major aspects: i) it
is proved that in center of mass frame and in internal coordinate
Hilbert subspace, the total angular momentum $J$ of meson is
conserved and the mass eigen equation can be expressed in total
angular momentum representation and in terms of a set of coupled
radial eigen equations for each $J$; ii) Based on lattice QCD
results, a relativistic confining potential is introduced into the
effective interaction and the excited states of mesons can be well
described; iii) an SU(3) flavor mixing interaction is introduced
phenomenologically to describe the flavor mixing mesons and the mass
eigen equations contain the coupling among different flavor
components; iv) the mass eigen equations are of relativistic
covariance and the coupled radial mass eigen equations take full
account of $L-S $ coupling and tensor interactions. The model has
been applied to describe the whole meson spectra of about 265 mesons
with available data, and the mass eigen equations have been solved
nonperturbatively and numerically. The agreement of the calculated
masses,  squared radii, and decay constants with data is quite well.
For the mesons whose mass data have large experimental uncertainty,
the model produces certain mass values for test. For some mesons
whose total angular momenta and parity are not assigned
experimentally, the model gives a prediction of the spectroscopic
configuration$^{2S+1}L_J$.  The connection between our model and the
recent low energy QCD issues-the infrared conformal scaling
invariance and holographic QCD hadron models is discussed.

\end{abstract}
\maketitle

\section{Introduction}

To study hadronic properties at low energy scales, nonperturbative
effects must be taken into account\cite{Wilson}. To describe mesons
and baryons, there are several main approaches: coupled
Bethe-Salpeter(BS) and Dyson-Schwinger(DS) equation approach,
relativistic constituent quark model based on Bethe-Salpeter
equation(BSE), relativistic string Hamiltonian approach, and
holographic light-front QCD approach. In the coupled Bethe-Salpeter
and Dyson-Schwinger equation approach by P. Maris, P. Tandy, L.
Kaptari et al.\cite {MarisTandy}, the dressed quark propagators are
assumed to have time like complex mass poles where the absence of
real mass poles simulates quark confinement; the BS kernel is
approximated by ladder rainbow truncation with two-parameter
infrared structure. The approach contains four parameters in u-d-s
quark sector and is consistent with quark and gluon confinement.
Besides, it has the feature of preserving the relevant Ward identity
and generating Dynamical chiral symmetry breaking. The vector mesons
$\rho, \phi$, and $ K^*$ are studied in detail, the calculated
masses of $\rho, \phi$,and $ K^*$ mesons and decay constants
$f_{\rho}, f_{\phi}$, and $f_{K^*}$ are within $5\%$ and $10\%$ of
the data respectively. Moreover, the ground-state spectra of
light-quark mesons are also studied and a good description of
flavor-octet pseudoscalar, vector, and axial-vector meson spectrum
is obtained. The applicable domain of ladder truncation and the
relative importance of various components of the two-body BS
amplitude are also explored. However£¬~heavy quark mesons are not
investigated and the number of mesons treated are not too many. R.
Alkofer, P. Watson, and H. Weigel \cite{AlkoferMeigel} follow the
same approach, scalar and pseudoscalar, vector and axial vector
mesons are studied. A similar approach is pursued by P. Jain and
Munczek\cite{Munczek}, about 50 mesons are investigated and the
results are in good agreement with experiments. But heavy quarks are
analyzed by non-relativistic dynamics. It should be noted that in
contrary to Hamiltonian dynamics which works with wave functions
that are not manifestly covariant quantities, the above BSE/DSE
approaches emphasize the relativistic covariant aspect of the
formalism and invariant quantities are studied.

The constituent quark model(CQM) works surprisingly well for most of
the observed hadronic states \cite{Isgur01,Hersbach01}. However, as
a phenomenological theory, there are still some problems and puzzles
that need to be clarified and understood\cite{Godfrey}. One of the
most important problems is relativistic effect. To solve the
relativistic covariant problem of CQM, the relativistic constituent
quark model based on Bethe-Salpeter equation was proposed by B.
Metsch et al. in Bonn Group \cite{Metsch}. In this approach, the
meson and baryon Hamiltonians are extracted from Bethe-Salpeter
equation and the relativistic covariant constituent quark models for
mesons and baryons are constructed. Based on Dirac structure of the
two-body effective interactions, two types of models( A and B) are
constructed. This approach addresses hadron mass spectra from ground
state to 3GeV, light-flavor mesons, scalar excitations, linear Regge
trajectory, pseudoscalar mixing, and parity doublet(for baryons). In
this approach, the Dyson-Schwinger equation(DSE) is approximated by
parametrization of infrared effective gluon propagator, the
interaction kernel of BSE is given by single gluon exchange(OGE) and
the confinement is parameterized by a string-like potential ( having
two versions defined by Dirac structures A and B ). The
instanton-induced spin-flavor dependent interaction is also included
in the BSE kernel. The mass spectra up to 3 GeV, electroweak and
strong-decay properties are calculated with 7 to 9 parameters. About
60  scalar and pseudo-scalar, vector and axial vector, and some
tensor mesons with J=0,1,2 are calculated by models A and B, and
compared to Godfrey-Isgur's calculation and experimental data ( the
deviation seems large but the errors are not indicated ). Due to the
Dirac structure of the effective interactions, spin-spin and
spin-orbital interactions are included. Besides, heavy mesons are
not treated.

The relativistic string Hamiltonian approach was proposed by A.M.
Badalian et al.\cite{Badalian}. The merit of this approach is that
the quark-anti-quark interaction and the confinement are generated
by the relativistic string ( through Nambu-Goto action for QCD
vacuum fluctuation) which leads to a large reduction of the number
of model parameters. After quantizing the action by path integral,
they construct a Hamiltonian with a linear confining potential and
hyperfine quark-anti-quark interactions. Using only one parameter of
string tension, they study the systematic property of orbital
excitations and rotation of mesons. The linear Regge trajectory
relation between squared mass and orbital angular momentum is
produced nicely and in agreement with the data for about 40 mesons.
The relativistic string Hamiltonian approach is spin-independent. In
the lowest order, this approach doesn't contain spin-spin,
spin-orbital, and tensor interactions, thus it can produce the spin
averaged mass spectra for mesons. However, to include the higher
order effects by perturbation method, the hyperfine spin-dependent
interactions could be obtained.

 The holographic light-front QCD approach by S. J. Brodsky and G. F. de
Teramond et al. \cite{Brod} is based on light-front QCD and AdS/CFT
correspondence. The AdS/CFT correspondence between string theory in
AdS space and conformal field theories in physical space-time leads
to an analytic, semi-classical model for strongly-coupled QCD, which
has scale invariance and dimensional counting at short distances and
color confinement at large distances. This correspondence also
provides AdS/CFT or holographic QCD predictions for the analytic
form of the frame-independent light-front wave functions (LFWFs) and
masses of mesons and baryons. Recently, Brodsky $et\ al.$\cite{Brod}
have found that the transverse separation of quarks within hadron is
related to holographic coordinate (the fifth dimensional
z-coordinate) in AdS/CFT correspondence, the mass eigen equation of
meson in light-front effective Hamiltonian approach corresponds to
the equation of motion for the holographic field of effective
gravity field of super string in AdS space at low energy limit.
Recently, they have modified the gravitation background by using a
positive-sign dilaton metric to generate confinement and break
conformal symmetry. In the meanwhile, the chiral symmetry is broken
and a mass scale is introduced to simulate the effect. Based on
AdS/CFT correspondence, the holographic light-front QCD model yields
a first order description of some hadronic spectra. This model is
quite appealing and promising, since it has established a profound
relationship between super string theory and QCD in low energy
limit. In this model, very few parameters (cutoff parameter
$\Lambda_{QCD}$) are used to obtain the spectra for both mesons and
baryons, such as $\pi$, $\rho$, and $\Delta$, etc., which fit the
experimental data well\cite{Brod}. However, for the large body of
mesons, only few of them are described properly and a large part of
mesons are still left over. Besides, in its preset form the full
spin interactions are not treated properly although it has potential
to describe spin splittings.

 The light-front formalism\cite{Dirac} provides a convenient
nonperturbative framework for the relativistic description of
hadrons in terms of quark and gluon degrees of freedom\cite{Ma}.
Some fundamental nonperturbative light-front QCD approaches are
available, such as light-front Bethe-Salpeter
approach\cite{Kisslinger}, holographic light-front QCD
model\cite{Brod}, and light-front Hamiltonian method\cite{BPP}. The
light-front Bethe-Salpeter approach has been proposed by Kisslinger
et al. to study pion form factor and the transition from
non-perturbative to perturbative QCD calculation of pion form
factor. Like the B-S approach of instant form, the equation of
motion for light-front B-S wave function should be solved together
with Schwinger-Dyson equation for dressed quark propagator, vertex,
and self-energy, and the model parameters include confining
potential strengths, and others for parametrization of the BS Kernel
and the running quark masses. An interesting conclusion drawn from
the study of this approach is that the perturbative QCD calculation
works at the energy of 4-5 GeV, much lower than that explored
previously by the instant form of QCD.

The effective light-front QCD Hamiltonian theory proposed by Brodsky
and Pauli \cite{BPP} is an attempt to describe the hadron structure
as a bound constituent quark system in terms of Fock-space for the
light-front wave-function. The effective Hamiltonian of the approach
has been constructed recursively from the larger valence quark and
anti-quark Fock sectors and reduced to the lowest valence
quark-anti-quark sector\cite{Zhou}. Because of some unique features,
particularly the apparent simplicity of the light-front vacuum, this
model is a promising approach to the bound-state problem of
relativistic composite systems. Within the framework of the
discretized light-front QCD, Pauli $et\, al.$ have derived
non-perturbatively an effective light-front Hamiltonian for mesons,
which acts only on the $q\bar{q}$ sector\cite{Pauli01,Pauli02}. The
mass eigen equations of mesons are formulated in momentum-helicity
representation which hinders its solution in total angular momentum
representation. Besides, in this effective Hamiltonian, confining
potentials and flavor mixing interactions are lacking, so that the
excited states of mesons and flavor diagonal light mesons can not be
treated properly\cite{Wang}.

In order to apply the approach to describe mesons in whole
$q\bar{q}$ sector, essential changes are needed. First we have
proved that in center of mass frame (rest frame) and in internal
coordinate Hilbert subspace, the total angular momentum of the meson
system is conserved(see Appendix A and B ). Then we are working in
center of mass frame and in internal coordinate Hilbert subspace and
make the following three significant improvements on the model: (1)
transforming  mass eigen equations from momentum-spin representation
to total angular momentum representation  and establishing a set of
coupled radial mass eigen equations for each total angular momentum
; (2) introducing a relativistic confining potential into the
effective meson interaction phenomenologically based on lattice QCD
results ; (3) including an SU(3) flavor-mixing interaction in the
model phenomenologically and obtaining a set of coupled radial eigen
equations for different flavor components. In having done above,
finally we  have a complete QCD inspired relativistic bound state
model for mesons on the whole $q\bar{q}$ sector. This model has been
applied to about 265 mesons with available data and with total
angular momentum from $J=0$ to $6$. The mass spectra, squared radii,
and decay constants are calculated, and the calculated results are
in good agreement with the data. While the most important physical
results have been reported briefly in a short letter\cite{PLB}, the
present article will provide detailed information and solid
foundation of the model for completion

This paper is organized as follows. In Sec. \Rmnum{2} the QCD
inspired relativistic bound state model for mesons is described and
the relativistic mass eigen equations for bound states with any
total angular momentum are derived. In Sec. \Rmnum{3} based on
lattice QCD results, a relativistic confining potential in momentum
space is introduced in the effective interaction of mesons. The
effective interaction is extended to include an SU(3) flavor mixing
interaction in Sec. \Rmnum{4}. In Sec. \Rmnum{5} we present the
numerical solutions for 265 mesons including both flavor-off and
flavor diagonal mesons with $J=0-6$. Sec.\Rmnum{6} is an analysis of
the results obtained. Finally, conclusion and discussion are given
in Sec. \Rmnum{7}. The four Appendices are for clarifying some
important issues and for the derivation of key equations.

\newpage

\section{ Description of the model}

  For convenience, Brodsky and Pauli defined a light-front Lorentz invariant
Hamiltonian \cite{BPP}
\begin{eqnarray}
H_{\mathrm{LC}}\equiv P^\mu P_\mu = P^- P^+ - {\bm P}^2_\perp
=\hat{M}_0^2 ,
\end{eqnarray}

The relativistic bound state problem in front form can be solved by
solving the light-front mass eigen equation:
\begin{eqnarray}
\label{HLCeq} H_{\mathrm{LC}}\vert \Psi \rangle=M_0^2 \vert \Psi
\rangle \ .
\end{eqnarray}

If one disregards possible zero modes and works in the light-front
gauge, this equation can be solved in terms of a complete set of
Fock states $\vert \mu_n \rangle$:
\begin{eqnarray}
\sum_{n^\prime}\!\int \! d[\mu^\prime_{n^\prime}] \langle \mu_n
\vert H_{\mathrm{LC}}\vert \mu^\prime_{n^\prime} \rangle\langle
\mu^\prime_{n^\prime} \vert \Psi \rangle=M_0^2 \langle \mu_n \vert
\Psi \rangle \ .
\end{eqnarray}
For a meson,  the ket $\vert \Psi \rangle$ holds:
\begin{eqnarray}
\lefteqn{ \vert \Psi_{\mathrm{meson}} \rangle  = \sum_i
\Psi_{q\bar{q}}(x_i,
\vec k _{\!\perp i},\lambda_i)\vert q\bar{q} \rangle } \nonumber\\
& &+ \sum_i \Psi_{gg}(x_i, \vec k _{\!\perp i},\lambda_i)\vert
gg \rangle\nonumber\\
& &+ \sum_i \Psi_{q\bar{q}g}(x_i, \vec k _{\!\perp
i},\lambda_i)\vert q\bar{q}g \rangle\nonumber\\
& &+ \sum_i \Psi_{q\bar{q}q\bar{q}}(x_i, \vec k _{\!\perp
i},\lambda_i)\vert
q\bar{q}q\bar{q} \rangle\nonumber\\
& &+...
\end{eqnarray}

Within the framework of discrete quantization of light-front QCD,
infinite dimensional Fock  space has been truncated at a proper
cutoff energy and the energy truncation plays a role of
   renormalization in discrete light-front QCD. By Tamm-Dancoff projection method
 and resolvent technique, the equation of motion in a larger Fock space of multi-particles
 can be reduced to that in a smaller one with an effective interaction to account for
 the effect of the projected out part of the Fock space. The reduction and projection procedure
 can be carried out recursively, finally the effective Hamiltonian and its eigen equation on
 $q\bar{q}$ sector  can be obtained. For flavor off-diagonal mesons, disregarding the
 zero modes and the two-gluon annihilation effect, Pauli et al. has obtained the
effective mass eigen equation for mesons in light-front relative
momentum coordinate space \cite{Pauli01, Pauli02}
\begin{eqnarray}
\label{eq:pauli} \lefteqn{
    M_0^2\langle x,\vec k_{\!\perp}; \lambda_{q},
    \lambda_{\bar q}  \vert \psi\rangle =
}\nonumber\\ & & \left[
    {\overline m^{\,2}_{q} + \vec k_{\!\perp}^{\,2}\over x } +
    {\overline m^{\,2}_{\bar  q} + \vec k_{\!\perp}^{\,2}\over 1-x }
    \right]\langle x,\vec k_{\!\perp}; \lambda_{q},
    \lambda_{\bar q}  \vert \psi\rangle
\nonumber\\ & &
     - {4\over 3}{m_1m_2 \over \pi^2}
    \sum _{ \lambda_q^\prime,\lambda_{\bar q}^\prime}
    \!\int\!\frac{dx^\prime d^2 \vec k_{\!\perp}^\prime
    \,R(x^\prime,k_{\perp}^\prime)}
    {\sqrt{x(1-x)x^\prime(1-x^\prime)}}
\nonumber\\ & &
    {\overline\alpha(Q) \over Q  ^2} \,
    S_{\lambda_q\lambda_{\bar q};\lambda_q^\prime
    \lambda_{\bar q}^\prime}
    \,\langle x^\prime,\vec k_{\!\perp}^\prime;
    \lambda_q^\prime,\lambda_{\bar q}^\prime
    \vert \psi\rangle \ ,
\end {eqnarray}

 This is a relativistic covariant mass eigen
equation for mesons in center of mass fame and in internal Hilbert
subspace. However the equation of motion is written in relative
momentum and helicity representation, and the momentum-helicity
plane wave function contains all possible components of partial
waves of the spin spherical harmonic functions $\Phi_{JlsM}$, the
total angular momentum $J$ and its $z$-component $M$ are not
conserved.

Despite this, Trittmann and Pauli\cite{Pauli03}  found an
appropriate method which can calculate the eigenvalue spectrum
separately for each $ J_z=M$. To do so, they transformed the
light-front coordinate $x$ back to the coordinate $k_3$ by Terent'ev
transformation\cite{Tere}, and used a unitary transformation to
transform the Lepage-Brodsky spinors to the Bjorken-Drell
spinors\cite{Pauli04}. Then the mass eigen equation \eqref{eq:pauli}
becomes\cite{Pauli05}:
\begin{eqnarray}
\label{eq:pauli1}
 \lefteqn{ \hspace*{-1cm}
  \left[ M_0^{2} - \left( E_{1}(k)+E_{2}(k) \right)^{2}
  \right] \varphi_{s_{1}s_{2}}(\bm{k})
 }
 \nonumber \\
 & = &
 \sum_{s_{1}^{\prime}s_{2}^{\prime}} \int d^{3}\bm{k'}
 U_{s_{1}s_{2};s_{1}^{\prime}s_{2}^{\prime}}(\bm{k};\bm{k^\prime})
 \varphi_{s_{1}^{\prime }s_{2}^{\prime}}(\bm{k}^{\prime}),
 \label{eq:main0}
\end{eqnarray}

This integration equation is written in momentum-spin representation
in terms of internal relative momenta of two quarks, spin singlet
and triplet are mixed. For the same reason as discussed above, the
momentum-spin plane wave does not conserve $J$ and $M$. As noted in
Ref\cite{BPP}, in general it is difficult to explicitly compute the
total angular momentum of a bound state by using light-front
quantization. However, as addressed in Introduction, in the center
of mass frame and in internal Hilbert subspace, the total angular
momentum is conserved. This makes it possible to solve the mass
eigen equation in total angular momentum representation (see
Appendices B,C,D ).

Since  in center of mass frame and in internal Hilbert subspace, the
total angular momentum $J^2$ and $J_z$ are conserved, we can
transform the mass eigen equation (6) from momentum-spin
representation to total angular momentum representation and
establish the mass eigen equation for each $J$. Expanding the
momentum-spin plane wave function in terms of the spin spherical
harmonic functions $\Phi_{JslM}(\Omega_{k},s_1,s_2)$ and projecting
out the spin and angular part of the wave function in $|JslM\rangle$
subspace by the projecting operation,
\begin{equation}
 \Big\langle
  \sum_{m\mu}\sum_{s_{1}s_{2}} \langle lms\mu | JM \rangle
   \langle \frac{1}{2}s_{1}\frac{1}{2}s_{2} |s\mu \rangle
   Y_{lm}(\Omega_{k})\chi(s_1),\chi(s_2)
 \Big| ,
\end{equation}
 we obtain the mass eigen equation  for the radial
wave function of $R_{Jsl}(k)$ (see Appendix C).
\begin{eqnarray}
\label{eq:m}
 \lefteqn{
  \left[ M_0^{2}-\left( E_{1}(k)+E_{2}(k)\right) ^{2}\right] R_{Jsl}(k)
 }
 \\
 & = &
  \sum_{l^{\prime}=|J-s'|}^{J+s'} \sum_{s'=0,1} \int k^{\prime 2}dk^{\prime}
   U_{sl;s^{\prime} l^{\prime}}^{J}(k;k^{\prime})
   R_{Js'l^{\prime}}(k^{\prime}) .
 \nonumber
\label{eq:wang}
\end{eqnarray}
This is a set of coupled equations for radial functions $R_{Jsl}(k)$
of different partial waves and of spin singlet and triplets, coupled
by the tensor potential and by the relativistic spin-orbital
potential. In this case, the eigen wave functions $R_{Jsl}(k)$ has
the conventional definition and physical meaning. The bound states
of mesons can be described concisely by the spectroscopic symbol of
$^{2S+1}L_J$.

The kernel $U_{sl;s^{\prime}l^{\prime}}^{J}(k;k^{\prime})$ can be
written as (see Appendix D ),
\begin{eqnarray}
 \lefteqn{
  U_{sl;s^{\prime}l^{\prime}}^{J}(k;k^{\prime})
  =
  \sum_{mm^{\prime}} \sum_{s_{1}s_{2}} \sum_{s_{1}^{\prime}s_{2}^{\prime}}
  \int \int d\Omega_{k} d\Omega_{k^{\prime}}
 }
 \nonumber \\
 &   & \hspace*{-0.2cm} \mbox{}\times
 \langle Y_{lm}(\Omega_{k}) |
         U_{s_{1}s_{2};s_{1}^{\prime}s_{2}^{\prime}}(\bm k,\bm k^{\prime})
         | Y_{l^{\prime}m^{\prime}}(\Omega_{k^{\prime}})
 \rangle
 \\
 &   & \hspace*{-0.2cm} \mbox{} \times
 \langle lms\mu |JM \rangle
 \textstyle\langle \frac{1}{2}s_{1}\frac{1}{2}s_{2} | s\mu\rangle
 \langle l^{\prime}m^{\prime}s^{\prime}\mu^{\prime} | JM\rangle
 \textstyle\langle \frac{1}{2}s_{1}^{\prime}\frac{1}{2}s_{2}^{\prime} |
                   s^{\prime}\mu^{\prime}
           \rangle .
 \nonumber
\end{eqnarray}

The above kernel $U_{sl;s^{\prime}l^{\prime}}^{J}(k;k^{\prime})$
contains different kinds of central potentials, relativistic
spin-orbit coupling potentials, and tensor potentials changing $l$
by $\Delta l = \pm 2$ and mixing spin singlet and triplets (see
Appendix D ).

\section{Introducing a Confining potential}
Quark confinement is one of the fundamental problems in QCD for
hadronic physics. The confinement  and the spontaneous breaking of
chiral symmetry are key ingredients for solving the low-energy
hadronic bound states from QCD, but none of them has been completely
understood and solved. Numerical results show that the effective
light-front Hamiltonian model proposed by Pauli et al. without
confining potentials can well describe the ground states but can not
apply to the radial excited states of mesons.  To describe the
excited states properly, the confining potential must be included in
the model\cite{Wang}.

  Fortunately, we can refer to the constituent quark model which is
successful due to the inclusion of a phenomenological confining
potential in some way\cite{LQCD}. The key idea of this model
consists in the introduction of a linear confining potential in
coordinate space based on the numerical calculations of lattice QCD,
and this non-relativistic confining potential can be generalized to
relativistic form.

In nonrelativistic quark models the confining potential in
configuration space is,
\begin{eqnarray}
\label{NRcon} V_{\mathrm{con}}(r)=\lambda r+c \ ,
\end{eqnarray}
where $\lambda$ is the strength of the linear interaction, and $c$
is a constant irrelevant in the present case and omitted hereafter.
By Fourier transformation, the counterpart of the linear term
$\lambda r$ in momentum space is obtained,
\begin{eqnarray}
& & V_{\mathrm{lin}}(\bm q) \sim -\frac{1} { |{\bm q}|^4} \ , \nonumber\\
& &    \bm q=\bm k- \bm k^\prime \ .
\end{eqnarray}
At the point of $\bm q = 0$, the singularity indicates that the
directly transformed result of linear potential could not be
described correctly in momentum space, which results in an
ill-defined bound state equation\cite{Gross}. However, some
different methods were employed to solve this problem for the
relativistic case. In the present paper, the correct form for
$V_{\mathrm{lin}}(\bm q)$ is constructed by introducing a small
parameter $\eta$:
\begin{eqnarray}
\label{NRlin} V( \bm q)&=& \lim_{\eta \rightarrow 0}{\lambda \over
2\pi^2} \frac{\partial^{2}}{\partial \!\eta^{2}} \left[ {1 \over
|\bm q|^2+\eta^2} \right]
\end{eqnarray}

The relativistic linear potential in momentum space
$V_{\mathrm{lin}}(Q)$  is a direct generalization of the
nonrelativistic one, just replacing the nonrelativistic  $|\bm q|^2$
in (\ref{NRlin}) by the relativistic $Q^2$, which has the following
specification in \cite{Sommerer01,Hersbach01},
\begin{eqnarray}
Q^2=(\mathbf{k}-\mathbf{k'})^2+\varpi^2
\end{eqnarray}
and
\begin{eqnarray}
\varpi^2=(E_1-E_1')(E_2-E_2')
\end{eqnarray}
Then the form of relativistic confining potential is,
\begin{eqnarray}
\label{Recon} V_{\mathrm{con}}(Q)&=& \lim_{\eta \rightarrow
0}{\lambda \over 2 \pi^2} \frac{\partial^{2}}{\partial \!\eta^{2}}
\left[ {1 \over Q^2+\eta^2} \right]
\end{eqnarray}
Obviously, this confining potential is Lorentz covariant and can be
used in either spin system or non-spin system.

Now as the relativistic confining potential $V_{\mathrm{con}}(Q)$ is
included in the interaction, one has the new kernel
$U_{s_1s_2;s^{\prime}_1 s^{\prime}_2}(\textbf{k},
\textbf{k}^{\prime})$,
\begin{eqnarray}
\label{eq:kernel}
& &U_{s_{1}s_{2};s_{1}^{\prime}s_{2}^{\prime}}(\textbf{k}, \textbf{k}^{\prime})=\\
& & \frac{4}{3}\frac{m_1m_2}{\pi^2}\sqrt{(\frac{1}{E_1}\!
+\!\frac{1}{E_2}) (\frac{1}{E_1^\prime}\!+\!\frac{1}{E_2^\prime})}
\bar{u}(\mathbf{k},s_1) \bar{u}(-\mathbf{k},s_2) \nonumber \\
& & \times  [\gamma^{(1)}_\mu \cdot \gamma^{(2)
\mu}V_V+\mathbb{I}^{(1)} \cdot \mathbb{I}^{(2)}V_S]
u(\mathbf{k^\prime},s_1^\prime)u(-\mathbf{k^\prime},s_2^\prime)  .
\end{eqnarray}
The scalar and vector interaction potentials read
\begin{eqnarray}
V_V&=&-\frac{\bar{\alpha}(Q)}{Q^2}-\frac{3}{4}\epsilon\:
V_{\mathrm{con}}(Q) \nonumber\\
V_S&=&-\frac{3}{4}(1-\epsilon) V_{\mathrm{con}}(Q) \,
\end{eqnarray}
where $\epsilon$ represents the scalar-vector mixing of the
confining potential.

\section{Including a flavor mixing interaction }

It is extremely difficult to derive a simple form of  flavor mixing
interaction in the above effective Hamiltonian from light-front QCD
at present. However, without flavor mixing potential, one can not
deal with the flavor diagonal mesons such as $\pi^0$, $\rho^0$, and
$f_0$, etc. In the fundamental hadronic theory, the quarks of u, d,
and s have an approximate $SU(3)$ symmetry. Due to this symmetry,
the quarks fields transform each other under the $SU(3)$
transformation \cite{Weinberg},
\begin{displaymath}
\left( \begin{array}{c}
u  \\
d  \\
s
\end{array} \right) \longrightarrow
exp\  [i\sum_a(\theta_a^V T_a+\theta_a^A T_a\gamma_5)]
 \left(
\begin{array}{c}
u  \\
d  \\
s
\end{array} \right)
\end{displaymath}
where $T_a$ are Gell-Mann Matrices. For convenience of numerical
calculation, we introduce phenomenologically a simple flavor mixing
interaction as follows,
\begin{align}
V_f=&\gamma_0\left[T^{+}_{ud}(1)T^{+}_{ud}(2)+T^{-}_{ud}(1)T^{-}_{ud}(2)\right]\nonumber\\
+&\delta_0\left[T^{+}_{us}(1)T^{+}_{us}(2)+T^{-}_{us}(1)T^{-}_{us}(2)\right.\nonumber\\
+&\left.T^{+}_{ds}(1)T^{+}_{ds}(2)+T^{-}_{ds}(1)T^{-}_{ds}(2)\right]
\end{align}
where $\gamma_0$ and $\delta_0$ are the strengths of flavor-mixing
interaction, the index $1$ and $2$ denote the quark and
 anti-quark in meson, respectively. The flavor SU(3) wave functions and generators
 are defined as,
\begin{align}
&|u\rangle=\left(
            \begin{array}{c}
              1 \\
              0 \\
              0 \\
            \end{array}
          \right),
|\bar{u}\rangle=\left(
            \begin{array}{c}
              1 \\
              0 \\
              0 \\
            \end{array}
          \right);
|d\rangle=\left(
            \begin{array}{c}
              0 \\
              1 \\
              0 \\
           \end{array}
          \right)\\
&|\bar{d}\rangle=\left(
            \begin{array}{c}
              0 \\
              1 \\
              0 \\
            \end{array}
          \right),
|s\rangle=\left(
            \begin{array}{c}
              0 \\
              0 \\
              1 \\
           \end{array}
          \right)
|\bar{s}\rangle=\left(
            \begin{array}{c}
              0 \\
              0 \\
              1 \\
            \end{array}
          \right)\\
&T_{ud}^{+}=(T_{ud}^{-})^\dag=\left(
            \begin{array}{ccc}
              0 & 1 & 0 \\
              0 & 0 & 0 \\
              0 & 0 & 0 \\
            \end{array}
          \right),
\\
&T_{us}^{+}=(T_{us}^{-})^\dag=\left(
            \begin{array}{ccc}
              0 & 0 & 1 \\
              0 & 0 & 0 \\
              0 & 0 & 0 \\
            \end{array}
          \right)\\
&T_{ds}^{+}=(T_{ds}^{-})^\dag=\left(
            \begin{array}{ccc}
              0 & 0 & 0 \\
              0 & 0 & 1 \\
              0 & 0 & 0 \\
            \end{array}
          \right),
\end{align}

The action of the flavor mixing interaction on flavor wave function
is as follows,
\begin{align}
V_f|u\bar{u}\rangle&=\gamma_0|d\bar{d}\rangle+\delta_0|s\bar{s}\rangle\\
V_f|d\bar{d}\rangle&=\gamma_0|u\bar{u}\rangle+\delta_0|s\bar{s}\rangle\\
V_f|s\bar{s}\rangle&=\delta_0|d\bar{d}\rangle+\delta_0|u\bar{u}\rangle
\end{align}

Combining this interaction  with the precious one in equations (18)
, we have a set of flavor-coupled  radial eigen equations for the
flavor components of  up, down, and strange quarks,
\begin{eqnarray}
 \lefteqn{
  \left[ M_0^{2}-\left( E_{1}(k)+E_{2}(k)\right) ^{2}\right] R_{Jsl}^{p_1p_2}(k)
 }
 \\
 & = &
  \sum_{l^{\prime}=|J-s'|}^{J+s'} \sum_{s'=0,1} \sum_{p'_1p'_2} \int k^{\prime 2}dk^{\prime}
   U_{Jsl;s^{\prime} l^{\prime}}^{p_1p_2,p'_1p'_2}(k;k^{\prime})
   R_{Js'l^{\prime}}^{p'_1p'_2}(k^{\prime}) .
 \nonumber
\label{eq:wangf}
\end{eqnarray}
The interaction kernel including the flavor-mixing interaction is
\begin{align}
&U_{Jsl;s'l'}^{p_1p_2;p'_1p'_2}(k,k')=\sum_{mm'}\sum_{\mu\mu'}\sum_{s_1s_2}\sum_{s'_1s'_2}\int\int{d}
\Omega_k\Omega_{k'}\nonumber\\
&\times{Y^*_{lm}(\Omega_k)}W^{p_1p_2;p'_1p'_2}_{s_1s_2;s_1's_2'}(\textbf{k},\textbf{k}')Y_{l'm'}(\Omega_{k'})\\
&\langle{lms\mu}|JM\rangle\langle{\frac{1}{2}s_1\frac{1}{2}s_2}|s\mu\rangle
\langle{l'm's'\mu'}|JM\rangle\langle{\frac{1}{2}s'_1\frac{1}{2}s'_2}|s'\mu'\rangle\nonumber,
\end{align}
where
$W^{p_1p_2;p'_1p'_2}_{s_1s_2;s_1's_2'}(\textbf{k},\textbf{k}')$ is
defined as
\begin{align}
&W^{p_1p_2;p'_1p'_2}_{s_1s_2;s_1's_2'}(\textbf{k},\textbf{k}')=
\frac{4}{3}\frac{m_{p_1}m_{p_2}}{\pi^2}\sqrt{\left(\frac{1}{E_1}+\frac{1}{E_2}\right)
\left(\frac{1}{E'_1}+\frac{1}{E'_2}\right)}\times\nonumber\\
&{\bar{u}(p_1,\textbf{k},s_1)}\bar{u}(p_2,-\textbf{k},s_2)
\left[\gamma^{\mu}(p_1)\cdot\gamma_{\mu}(p_2)V_{V}
+I^{(1)}\cdot{I}^{(2)}V_S\right]\times\nonumber\\
&\left[I_{f}+V_f\right]
{{u}'(p'_1,\textbf{k}',s'_1)}{u}'(p'_2,-\textbf{k}',s'_2).
\end{align}
where $p_1p_2,p'_1p'_2=\{u\bar{u},d\bar{d},s\bar{s}\}$, $I_{f}$ is
the identity operator in flavor space, $V_V$ and $V_S$ are the
vector potential and scalar potential, respectively. For the flavor
mixing mesons with total angular momentum $J$,  the mass eigen
equations are described explicitly by the following set of
flavor-coupled equations
\begin{subequations}
\begin{align}
&\left[M_0^2-\left(E_u(k)+E_{\bar{u}}(k)\right)^2\right]R^{u\bar{u}}_{Jsl}(k)\label{k10}\\
&=\sum_{l'=|J-s'|}^{J+s'}\sum_{s'=0,1}\int{k'^2}dk'\left(U_{Jsl;s'l'}^{u\bar{u};
u\bar{u}}(k,k')R^{u\bar{u}}_{Js'l'}(k,k')
\right.+\nonumber\\
&\left.\gamma_0
U_{Jsl;s'l'}^{u\bar{u};d\bar{d}}(k,k')R^{d\bar{d}}_{Js'l'}(k,k')
+\delta_0
U_{Jsl;s'l'}^{u\bar{u};s\bar{s}}(k,k')R^{s\bar{s}}_{Js'l'}(k,k')
\right)\nonumber\\
&\left[M_0^2-\left(E_d(k)+E_{\bar{d}}(k)\right)^2\right]R^{d\bar{d}}_{Jsl}(k)\label{k11}\\
&=\sum_{l'=|J-s'|}^{J+s'}\sum_{s'=0,1}\int{k'^2}dk'\left(\gamma_0
U_{Jsl;s'l'}^{d\bar{d}; u\bar{u}}(k,k')R^{u\bar{u}}_{Js'l'}(k,k')
\right.+\nonumber\\
&\left.U_{Jsl;s'l'}^{d\bar{d};d\bar{d}}(k,k')R^{d\bar{d}}_{Js'l'}(k,k')+
\delta_0
U_{Jsl;s'l'}^{d\bar{d};s\bar{s}}(k,k')R^{s\bar{s}}_{Js'l'}(k,k')
\right)\nonumber\\
&\left[M_0^2-\left(E_s(k)+E_{\bar{s}}(k)\right)^2\right]R^{s\bar{s}}_{Jsl}(k)\label{k12}\\
&=\sum_{l'=|J-s'|}^{J+s'}\sum_{s'=0,1}\int{k'^2}dk'\left(\delta_0
U_{Jsl;s'l'}^{s\bar{s}; u\bar{u}}(k,k')R^{u\bar{u}}_{Js'l'}(k,k')
\right.+\nonumber\\
&\left. \delta_0
U_{Jsl;s'l'}^{s\bar{s};d\bar{d}}(k,k')R^{d\bar{d}}_{Js'l'}(k,k')+
U_{Jsl;s'l'}^{s\bar{s};s\bar{s}}(k,k')R^{s\bar{s}}_{Js'l'}(k,k')
\right)\nonumber
\end{align}
\end{subequations}

\section{Numerical Solutions}
In the above equations, $J, s, l$ denote total angular momentum,
total spin, and total orbital angular momentum, respectively.
Different mesons can be classified by the spectroscopic symbol
$^{2S+1}L_J$(or their combination), which is equivalent to the
symbol $J^{PC}$. The space parity and charge conjugation parity are
denoted as,
\begin{eqnarray}
P&=&(-1)^{L+1}\nonumber\\
C&=&(-1)^{L+S} .
\end{eqnarray}

In the present model, the mass eigen value problem of mesons is
described by a set of coupled integration equations, and the
interaction includes a quark-anti-quark one gluon exchange potential
$V_{OGE}$ , a confining potential $V_{con}$, and a flavor mixing
interaction $V_{f}$. If the confining potential has a pure
iso-scalar structure( $\epsilon \!=\! 0$) and the flavor mixing
interaction is omitted, the interaction contains two parameters: the
effective coupling constant $\bar{\alpha}$ and the confining
potential strength $\lambda$. Besides, the flavor mixing interaction
has two parameters, and the constituent quark masses are also
indispensable parameters to describe spontaneous chiral symmetry
breaking.

 The numerical solution of the eigen equations can be obtained
 by discretization of integration equation $(8)$ or $(31a-c)$, and
 the integration equations are transformed into matrix equations.
 The 4-fold integral of the kennel is completed by the integration technique of
 spherical harmonic functions and the angular momentum algebra, while the
 integration over $k$ is performed by  using Gauss-Legendre
 quadratures. The integration region $k\in [0,\infty)$ is projected
 onto the finite interval $x\in [-1,1]$ by $ x\!=\!\frac{k-1}{k+1} $.
 The radial mass eigen  equation (8) is discretized as follows:
\begin{eqnarray}
\label{eq:discrete}
 \lefteqn{
  \left[ M_0^{2}-\big( E_{1}(k_i)+E_{2}(k_i)
  \big) ^{2}\right] R_{Jsl}(k_i)
 }
 \\
 & = &
  \sum_{l^{\prime}=|J-s'|}^{J+s'} \sum_{s'=0,1} \sum_{j=1}^{N}
      U_{sl;s^{\prime} l^{\prime}}^{J}(k_i;k_j)
   R_{Js'l^{\prime}}(k_j) k_j^2 w_j \ ,
 \nonumber
\end{eqnarray}
where $w_j$ is the weight of integration. Before diagonalizing this
matrix equations,  special care should be taken for two kinds of
singularities: the singularity at infinite $k$ and the singularity
as $k=k^{\prime}$ inside the region of integration. The first one
has been solved by the projection of the region $k\in [0,\infty)$
 onto the finite interval $x\in [-1,1]$, and the second one is
treated by infrared singularity treatment. The detailed procedures
of calculation can be found in Ref.\cite{Numer,Wang}.

The parameters of the model are determined from best fit to
experimental data. In this paper, a purely scalar confining
potential($\epsilon=0$) is used. Reproducing the masses of
$\pi^{0}$, $\pi^{\pm}$, and $\pi(1300)$, we can determine
$\bar{\alpha}$, $\lambda$, and the masses of up and down quarks.
Then by reproducing the masses of $K^\pm$, $D^0$, and $B^\pm$, the
mass parameters of strange, charm, and bottom quarks are obtained.
The parameters of flavor mixing interaction are determined by the
best fit to the data of flavor diagonal mesons. From all the
available data of mesons \cite{Data} with $J=0-6$ ( 12 mesons are
left for future study: including 6 exotic mesons and 6 mesons
without any information about their $J,s,L$ ), we have obtained an
appropriate set of 6 parameters for flavor off-diagonal mesons:
$\bar{\alpha}=0.2574$, $\lambda=0.92\times 10^4 MeV^2$,
$m_{u/d}=0.297GeV, m_s=0.418GeV, m_c=1.353GeV, m_b=4.447GeV;$ and
for the flavor diagonal mesons: $\gamma_0=0.1$ and $\delta_0=0.1$.
The number of the model parameters is minimum for this kind of
semi-phenomenological models and comparable to BSE and CQM. The
masses and wave functions of scalar and pseudoscalar, vector and
axial-vector, tensor and pseudotensor mesons, and others with
$J=3-6$ have been calculated and compared with the experimental data
in the Table( including 265 mesons and anti-mesons: 123 (u,d)-light
mesons,  50 (s,u/d)-K mesons, 24 (c,u/d)-D mesons, 14 (s,c)-D
mesons, 12 (b,u/d)-B mesons, 10 (s,b)-$B_s$ mesons, 2 (c,b)-$B_c$
mesons, 16 (c,$\bar{c}$)
 mesons, and 14 (b,$\bar{b}$ ) mesons ). It is remarkable that among
265 mesons, 259 mesons are well described by this model within mass
error less than $23\%$.

 In addition, the radial wave-functions of mesons
in configuration space can be obtained from the radial
wave-functions in momentum space by Fourier transformation(see
Appendix C), then one can calculate the mean square radii and the
decay constants for some pseudoscalar mesons listed in
Tab.\ref{radii} and compared with experimental data.
\begin{table}[!htbp]
\renewcommand{\arraystretch}{1.2}
\caption{\label{radii} The mean square radii and decay constants of
some pseudoscalar mesons, compared  with the experimental
data\cite{Data}. (Radii are given in $fm^2$ and decay constants are
given in $MeV$)}
\begin{tabular}{ccccccc}
\hline
 & $\quad\pi^+\quad$ & $\quad K^+\quad$ & $\quad D^+\quad$ & $\quad D_s\quad$ & $\quad B\quad$ \\
\hline
$\langle r^2\rangle_{the}$      & 0.385  & 0.253 & 0.235 &-&-\\
$\langle r^2\rangle_{exp}$ & 0.452  & 0.314 & -&-& - \\
$f_{the}$    & 135.2   & 210.7   & 189.2 &253.1&227.5  \\
$f_{exp}$ & 130.4 & 155.5   & 205.8 &273&216   \\
\hline
\end{tabular}
\end{table}

\begin{longtable}[!htbp]{ccccc}
\caption{\label{pseudoscalar} The pseudoscalar mesons mass spectra
(in MeV). }
\endfirsthead
\hline Meson      & $I^G(J^{PC})$  & Exp(Mev) & Our's(Mev) & err(\%)\\
\hline
$\pi^0$           & $0^{-+}$  & 135              & 135     & 0  \\
$\pi^\pm$         & $0^{-+}$   & 140              & 140     & 0   \\
$\eta$            & $0^{-+}$  & 548              & 143     & 73   \\
$\eta(958)$       & $0^{-+}$  & 958              & 690     & 27 \\
$\eta(1295)$      & $0^{-+}$  & 1294             & 1258    & 2.8 \\
$\pi(1300)^\pm$   & $0^{-+}$   & 1300$\pm$100     &1408     & 0   \\
$\pi(1300)^0$     & $0^{-+}$  & 1300$\pm$100     & 1350    & 0 \\
$\eta(1405)$      & $0^{-+}$  & 1410           & 1652    & 17 \\
$\eta(1475)$      & $0^{-+}$  & 1476             & 1700    & 15 \\
$\eta(1760)$      & $0^{-+}$  & 1756             & 1769    & 0.8 \\
$\pi(1800)^\pm$   & $0^{-+}$   & 1816             & 1454    & 19 \\
$\pi(1800)^0$     & $0^{-+}$  & 1816             & 2096    & 15.4 \\
$X(1835)$         & $0^{-+}?$  & 1833             & 2110    & 15.1 \\
$\eta(2225)$      & $0^{-+}$  & 2220             & 2160    & 2.7 \\
$K^\pm$           & $0^{-}$   & 494              & 494     & 0   \\
$K^0$             & $0^{-}$   & 498              & 494     & 0.8 \\
$K(1460)$         & $0^{-}$   & 1460             & 1522    & 4.2 \\
$K(1830)$         & $0^{-}$   & 1830             & 1597    & 12.7\\
$D^0$             & $0^{-}$   & 1865             & 1931    & 3.5 \\
$D^\pm$           & $0^{-}$   & 1869             & 1931    & 3.3 \\
$D_s^\pm$         & $0^{-}$   & 1969             & 2001    & 1.6 \\
$B^0$             & $0^{-}$   & 5279             & 5584    & 5.8 \\
$B^\pm$           & $0^{-}$   & 5279             & 5584    & 5.8 \\
$B_s^0$           & $0^{-}$   & 5367             & 5667    & 5.6 \\
$B_c^\pm$         & $0^{-}$   & 6286             & 6342    & 0.9 \\
$\eta_c(1S)$      & $0^{-+}$  & 2980             & 2980    & 0   \\
$\eta_c(2S)$      & $0^{-+}$  & 3637             & 3533    & 2.9 \\
$\eta_b(1S)$      & $0^{-+}$  & 9391             & 8800    & 6.3   \\
\hline
\end{longtable}

\begin{longtable}[!htbp]{ccccc}
\caption{\label{scalar} The scalar mesons mass spectra (in MeV). }
\endfirsthead
\hline Meson      & $J^{PC}$  & Exp(Mev) & Our's(Mev) & err(\%)\\
\hline
$f_0(600)$        & $0^{++}$  & 400-1200       & 736        & 0     \\
$f_0(980)$        & $0^{++}$  & 980            & 994        & 1.4    \\
$a_0(980)^0$      & $0^{++}$  & 985            & 1080       & 9.6  \\
$a_0(980)^\pm$    & $0^{++}$  & 985            & 930        & 5.6 \\
$f_0(1370)$       & $0^{++}$  & 1200-1500      & 1231       &  0 \\
$a_0(1450)^0$     & $0^{++}$  & 1474           & 1333       &  9.5\\
$a_0(1450)^\pm$   & $0^{++}$  & 1474           & 1457       & 1.1 \\
$f_0(1500)$       & $0^{++}$  & 1505           & 1522       & 1.1 \\
$f_0(1710)$       & $0^{++}$  & 1724           & 1568       & 9.0 \\
$f_0(2020)$       & $0^{++}$  & 1992           & 1606       & 19.4 \\
$f_0(2100)$       & $0^{++}$  & 2103           & 1989       & 5.4 \\
$f_0(2200)$       & $0^{++}$  & 2189           & 2026       & 7.4 \\
$f_0(2330)$       & $0^{++}$  & 2321           & 2052       &11.6  \\
$K_0^*(800)$      & $0^{+}$   & 672            & 731        & 8.8 \\
$K_0^*(1430)$     & $0^{+}$   & 1412           & 1535       & 8.7\\
$D_0^*(2400)^0$   & $0^{+}$   & 2352           & 2254       & 4.2 \\
$D_0^*(2400)^\pm$ & $0^{+}$   & 2403           & 2254       & 6.2 \\
$D_{s0}^*(2317)^\pm$& $0^{+}$   & 2317           & 2169       & 6.4 \\
$\chi_{c0}(1P)$   & $0^{++}$  & 3415           & 3352       & 1.8 \\
$\chi_{b0}(1P)$   & $0^{++}$  & 9860           & 9860       & 0   \\
$\chi_{b0}(2P)$   & $0^{++}$  & 10232          & 9990       & 2.3 \\
\hline
\end{longtable}

\begin{longtable}[!htbp]{ccccc}
\caption{\label{axialvector} The axial vector meson mass spectra (in
MeV).}
\endfirsthead
\hline
Meson      & $J^{PC}$  & Exp(Mev) & Our's(Mev) & err(\%)\\
\hline
$h_1(1170)$       & $1^{+-}$   & 1170     & 1027    & 12.2   \\
$b_1(1235)^0$     & $1^{+-}$   & 1230     & 1127    & 8.4   \\
$b_1(1235)^\pm$   & $1^{+-}$   & 1229     & 1343    & 9.3 \\
$a_1(1260)^0$     & $1^{++}$   & 1230     & 1276    & 3.7   \\
$a_1(1260)^\pm$   & $1^{++}$   & 1230     & 1371    & 11.4 \\
$f_1(1285)$       & $1^{++}$   & 1281     & 1295    & 1.1   \\
$h_1(1380)$       & $1^{+-}$   & 1386     & 1301    & 6.1   \\
$f_1(1420)$       & $1^{++}$   & 1426     & 1311    & 8.0   \\
$f_1(1510)$       & $1^{++}$   & 1518     & 1419    & 6.5   \\
$h_1(1595)$       & $1^{+-}$   & 1594     & 1495    & 6.2   \\
$a_1(1640)^0$     & $1^{++}$   & 1647     & 1745    & 6.0   \\
$a_1(1640)^\pm$   & $1^{++}$   & 1647     & 1724    & 4.7   \\
$K_1(1270)$       & $1^{+}$    & 1273     & 1459    & 14.6\\
$K_1(1400)$       & $1^{+}$    & 1402     & 1484    & 5.8 \\
$K_1(1650)$       & $1^{+}$    & 1650     & 1757    & 6.4 \\
$D_1(2420)^0$     & $1^{+}$    & 2422     & 2400    & 0.9 \\
$D_1(2420)^\pm$   & $1^{+}?$    & 2423     & 2400    & 0.9 \\
$D_1(2430)^0$     & $1^{+}$    & 2427     & 2425    & 0.1 \\
$D_{S1}(2460)^\pm$& $1^{+}$      & 2460     & 2530    & 2.9 \\
$D_{S1}(2536)^\pm$& $1^{+}$      & 2535     & 2549    & 0.6 \\
$B_{1}(5721)^0$   & $1^{+}$    & 5721     & 5666    & 1.0 \\
$B_{S1}(5830)^0$  & $1^{+}$    & 5829     & 5800    & 0.5 \\
$\chi_{c1}(1p)$   & $1^{++}$   & 3510     & 3504    & 0.2 \\
$h_{c1}(1p)$      & $1^{+-}$   & 3526     & 3509    & 0.5 \\
$\chi_{b1}(1p)$   & $1^{++}$   & 9892     & 10040   & 1.5 \\
$\chi_{b1}(2p)$   & $1^{++}$   & 10255    & 10040   & 2.1 \\
\hline
\end{longtable}

\begin{longtable}[!htbp]{ccccc}
\caption{\label{vector} The vector meson mass spectra (in MeV).}
\endfirsthead
\hline Meson      & $J^{PC}$ & Exp(Mev) & Our's(Mev) & err(\%)\\
\hline
$\rho(770)^0$     & $1^{--}$  & 775       & 1015    & 31    \\
$\rho(770)^\pm$   & $1^{--}$  & 775       & 1239    & 60\\
$\omega(782)$     & $1^{--}$  & 783       & 1270    & 62   \\
$\phi(1020)$      & $1^{--}$  & 1019      & 1334    & 31   \\
$\omega(1420)$    & $1^{--}$  & 1425      & 1410    & 1.0   \\
$\rho(1450)^0$    & $1^{--}$  & 1465      & 1636    & 11.6   \\
$\rho(1450)^\pm$  & $1^{--}$  & 1465      & 1323    & 9.7 \\
$\rho(1570)^0$    & $1^{--}$  & 1570      & 1641    & 4.5   \\
$\rho(1570)^\pm$  & $1^{--}$  & 1570      & 1740    & 10.8   \\
$\omega(1650)$    & $1^{--}$  & 1670      & 1675    & 0.3   \\
$\phi(1680)$      & $1^{--}$  & 1680      & 1786    & 6.3   \\
$\rho(1700)^0$    & $1^{--}$  & 1720      & 1836    & 6.7   \\
$\rho(1700)^\pm$  & $1^{--}$  & 1700      & 1362    & 19.8\\
$\rho(1900)^0$    & $1^{--}$  & 1909      & 1996    & 4.6   \\
$\rho(1900)^\pm$  & $1^{--}$  & 1909      & 1761    & 7.8   \\
$\rho(2150)^0$    & $1^{--}$  & 2149      & 2087    & 2.9   \\
$\rho(2150)^\pm$  & $1^{--}$  & 2149      & 2430    & 13.1 \\
$K^*(892)$        & $1^{-}$   & 892       & 1345    & 50.1\\
$K^*(1410)$       & $1^{-}$   & 1414      & 1415    & 0.1 \\
$K^*(1630)$       & $1^{-}?$   & 1629      & 1502    & 7.8\\
$K^*(1680)$       & $1^{-}$   & 1717      & 1531    & 10.8\\
$D^*(2007)^0$     & $1^{-}$   & 2007      & 2100    & 4.6 \\
$D^*(2010)^\pm$   & $1^{-}$   & 2010      & 2126    & 5.8 \\
$D^*(2640)$       & $1^{-}?$   & 2637      & 2403    & 8.9 \\
$D_S^{*\pm}$      & $1^{-}?$     & 2112      & 2214    & 4.8 \\
$D_{S1}(2700)^\pm$& $1^{-}$     & 2690      & 2233    & 16.9\\
$B^*$             & $1^{-}$   & 5325      & 5518    & 3.6 \\
$B_S^*$           & $1^{-}$     & 5413      & 5625    & 3.9 \\
$J/\psi$          & $1^{--}$   & 3097     & 3284    & 6.0 \\
$\psi(2S)$        & $1^{--}$   & 3686     & 3362    & 8.8 \\
$\psi(3770)$      & $1^{--}$   & 3773     & 3684    & 2.3 \\
$\psi(4040)$      & $1^{--}$   & 4039     & 3700    & 8.4 \\
$\psi(4160)$      & $1^{--}$   & 4153     & 4197    & 1.1 \\
$X(4260)$         & $1^{--}$   & 4263     & 4769    & 11.9 \\
$X(4360)$         & $1^{--}$   & 4361     & 4783    & 9.7 \\
$\psi(4415)$      & $1^{--}$   & 4421     & 5341    & 20.8 \\
$X(4660)$         & $1^{--}$   & 4664     & 5418    & 16.2 \\
$\gamma(1S)$      & $1^{--}$   & 9460     & 9693    & 2.4 \\
$\gamma(2S)$      & $1^{--}$   & 10023    & 10013   & 0.1 \\
$\gamma(3S)$      & $1^{--}$   & 10355    & 10060   & 2.8 \\
$\gamma(4S)$      & $1^{--}$   & 10580    & 10765   & 1.7 \\
$\gamma(10860)$   & $1^{--}$   & 10865    & 10783   & 0.8 \\
$\gamma(11020)$   & $1^{--}$   & 11019    & 10861   & 1.4 \\
\hline
\end{longtable}

\begin{longtable}[!htbp]{ccccc}
\caption{\label{tensor} The tensor and pseudotensor meson mass (in
MeV).}
\endfirsthead
\hline Meson & $J^{PC}$ & Exp(MeV) & Theor(MeV)& err(\%) \\
\hline
$\pi_2(1670)$     & $2^{-+}$  & 1672     & 1587    & 5.1\\
$\pi_2(1880)$     & $2^{-+}$  & 1895     & 1589    & 16.1 \\
$\pi_2(2100)$     & $2^{-+}$  & 2090     & 1922    & 8.0 \\
$K_2(1580)$       & $2^{-}$   & 1580     & 1530    & 3.2 \\
$K_2(1770)$       & $2^{-}$   & 1773     & 1539    & 13.2\\
$K_2(1820)$       & $2^{-}$   & 1816     & 1763    & 2.9 \\
$K_2(2250)$       & $2^{-}$   & 2247     & 1765    & 21.4\\
$\gamma(1D)$      & $2^{--}$  & 10161    & 10218   & 0.6 \\
$a_2(1320)$       & $2^{++}$  & 1318     & 1421    & 7.8 \\
$a_2(1700)$       & $2^{++}$  & 1723     & 1474    & 14.5\\
$K^*_2(1430)$     & $2^{+}$   & 1425     & 1531    & 7.4 \\
$K^*_2(1980)$     & $2^{+}$   & 1973     & 1575    & 20.1\\
$D^*_2(2460)^\pm$ & $2^{+}$   & 2460     & 2456    & 0.1 \\
$D^*_2(2460)^0$   & $2^{+}$   & 2462     & 2456    & 0.2 \\
$D^*_{S2}(2573)^\pm$&$2^{+}?$ & 2573     & 2580    & 0.3 \\
$B^*_J(5732)$     & $2^{+}?$   & 5698     & 5706     & 0.1 \\
$B^*_2(5747)^0$   & $2^{+}$   & 5743     & 5765    & 0.4 \\
$B^*_{S2}(5840)^0$& $2^{+}$   & 5840     & 5831    & 0.2 \\
$B^*_{SJ}(5850)^0$& $2^{+}?$     & 5853     & 5883    & 0.5 \\
$\chi_{c2}(1P)$   & $2^{++}$  & 3556     & 3732    & 4.9 \\
$\chi_{c2}(2P)$   & $2^{++}$  & 3929     & 3745    & 4.7 \\
$\chi_{b2}(1P)$   & $2^{++}$  & 9912     & 10014   & 1.0 \\
$\chi_{b2}(2P)$   & $2^{++}$  & 10269    & 10354   & 0.8 \\
\hline
\end{longtable}

\begin{longtable}[!htbp]{ccccc}
\caption{\label{tensor} The mesons of $J\geq3$ (in MeV).}
\endfirsthead
\hline Meson & $J^{PC}$ & Exp(MeV) & Theor(MeV)& err(\%) \\
\hline $\omega_3(1670)$  & $3^{--}$  & 1672     & 1677    & 0.3 \\
$\rho_3(1690)^\pm$& $3^{--}$  & 1688     & 1702    & 0.8 \\
$\rho_3(1690)^0$  & $3^{--}$  & 1688     & 1697    & 0.5 \\
$\phi_3(1850)$    & $3^{--}$  & 1854     & 1807    & 2.5 \\
$\rho_3(1990)^\pm$& $3^{--}$  & 1982     & 1795    & 9.4 \\
$\rho_3(1990)^0$  & $3^{--}$  & 1982     & 1807    & 8.8 \\
$\rho_3(2250)^\pm$& $3^{--}$  & 2230     & 2660    & 19.2\\
$\rho_3(2250)^0$  & $3^{--}$  & 2230     & 1852    & 17.0 \\
$K_3^*(1780)$     & $3^{-}$   & 2324     & 1777    & 23.5 \\
$K_3(2320)$       & $3^{+}$   & 2324     & 1812    & 22.0 \\
$a_4(2040)^\pm$   & $4^{++}$   &2001    & 1745     & 12.7  \\
$a_4(2040)^0$     & $4^{++}$   &2001    & 1743    & 12.9   \\
$f_4(2050)$       & $4^{++}$   &2018    & 1865     & 7.6   \\
$f_4(2300)$ & $4^{++}$   &2300    & 2016     & 12.3  \\
$K_4^*(2045)$     & $4^{+}$   & 2045     & 1827    & 10.6\\
$K_4(2500)$       & $4^{-}$   & 2490     & 1933    & 22.3 \\
$\rho_5(2350)^\pm$    & $5^{--}$   &2330     & 2292    & 1.6 \\
$\rho_5(2350)^0$    & $5^{--}$   &2330     & 2218     &  4.8  \\
$K_5^*(2380)$     & $5^{-}$   & 2382     & 2352    & 1.3 \\
$a_6(2450)^\pm$   & $6^{++}$&$2450\pm 130$ & 2412    & 0\\
$a_6(2450)^0$   & $6^{++}$&$2450\pm 130$ & 2423    & 0  \\
$f_6(2510)$       & $6^{++}$&$2465\pm 50$ & 2649    &  5.3 \\
\hline
\end{longtable}

\section{Analysis of the results }

In the above calculations, only one set of parameters are used,
which deserves discussion. The effective coupling strength or
running coupling constant $\bar{\alpha}$ and the related constituent
quark masses have a great influence on ground state of light mesons,
such as $\pi$. The confining potential strength $\lambda$ governs
the quark confinement at large distances and has strong influence on
the excited states of light mesons and also on the spectra of heavy
mesons. From the recent experiments of hadron physics, we know that
the QCD coupling $\alpha(Q^2)$ becomes large constant(not singular)
in the low momentum limit, which is called infrared conformal
invariance \cite{Conf}. This experimental fact explains why our
model with a set of constant parameters works well to describe the
structures of mesons in the energy region of
$0.14\text{GeV}\!\rightarrow\! 10\text{GeV}$, and our results may be
thought of confirming the infrared conformal invariance feature of
QCD on meson sector.

For light scalar mesons such as $a_0$ ,$K_0^*$, etc., although the
structure of the scalar mesons remains a challenging puzzle, our
model still describes $a_0(980)$, $a_0(1450)$, $K_0^*(800)$, etc.
quite well. For heavy mesons, because of the large masses of heavy
quarks, the effective double-gluon-exchange interactions for
off-diagonal heavy mesons are weak, which makes the model applicable
to them. Therefore, the calculated mass spectra for the mesons of
 $u/d\bar{s}$, $u/d\bar{c}$, $s\bar{c}$, $c\bar{c}$,
$c\bar{b}$, $u/d\bar{b}$, $s\bar{b}$, and $b\bar{b}$ are in good
agreement with the data. However the meson $K^{*}(892)$ on
$u/d\bar{s}$ sector with larger error of $50.1\%$ needs special
investigation(see below).

It should be noted that the $J$ and $P$ of $D_s^{*\pm}$ are not
identified by experiments, but their width and decay modes are
observed and consistent with the $1^-$ state. Nevertheless, our
model provides a definite assignment of $J=1$ and $P=-1$ for
$D_s^{*\pm}$. A similar prediction of the unidentified $J$ and $P$
is also made for  other 8 mesons: $X(1835)$, $D_1(2420)^\pm$,
$K^*(1630)$, $D^*(2640)$, $D^*_{S2}(2573)^\pm$,  $B^*_J(5732)$,
$B^*_{SJ}(5850)^0$, and $f_J(2220)$.

The 6 mesons  with  errors larger than $23\%$ provide some
information. For the vector mesons of $\eta,
\eta^{\prime}(985),\rho(770)^0$, $\phi(1020)$, and $\omega(782)$ on
$u/d$ sector, and $K^*(892)$ on $(u/d)s$ sector , the large
discrepancy indicates that the structures of these mesons are
special than others and need a different set of parameters: indeed,
as the set of parameters are re-adjusted to the set of
($\alpha=0.4594, \gamma_0=0.58, \delta_0=0.74$) and with the others
the same, a better fit is found with errors less than $23\%$.
Increase of the effective interaction strengths implies that these
vector mesons may have strong coupling between $q\bar{q}$ and
$qq\bar{q}\bar{q}$ subspaces and among different flavor components.

\section{Conclusion and discussion }

In conclusion, we have formulated the QCD inspired relativistic
bound state model for mesons and derived its mass eigen equations in
total angular momentum representation. It is proved that in center
of mass frame and in internal Hilbert subspace, total angular
momentum of the meson system is conserved. Moreover, by taking the
advantages of other effective QCD approaches
\cite{Sommerer01,Hersbach01}, the model has been improved
significantly by introducing both a relativistic confining potential
and an $SU(3)$ flavor mixing interaction. The resulting radial mass
eigen equations are solved numerically and nonperturbatively, and
265 mesons including flavor off-diagonal mesons and flavor diagonal
ones with $J=0-6$  are calculated and compared with experimental
data. The calculated masses are in good agreement with the data
within the mean square root mass error of $14\%$, only 6 mesons with
mass error larger than $23\%$. Besides, the wave functions obtained
from the model also yield reasonable mean square radii and decay
constants for some pseudo scalar mesons. In view that the structure
of the light scalar mesons is still a subject of
controversy\cite{Don}, and the internal dynamics of heavy-light
mesons in the static limit is far more complicated than that of the
heavy-heavy ones\cite{Dam}, our model can be thought to be
successful to describe  a large body of mesons.

  The comparison of our model with other approaches is as follows:

 1. As Pauli's model is concerned, we have improved the model
significantly on 5 important points and make it a predictive and
systematic model for mesons: 1) Proving that in internal Hilbert
subspace, total angular momentum is conserved; 2) establishing the
mass eigen equations in total angular representation for the first
time; 3) introducing the relativistic confining potential into the
model, which is new and quite different from Pauli, and its form
taken from the \cite{Sommerer01,Hersbach01}; 4) including the flavor
mixing interaction; 5) solving the mass equations for 265 mesons
nonperturbatively and the results are in good agreement with the
data.

 2. Comparing to other BSE and CQM meson models, our model is novel
in following points: 1) The effective Hamiltonian is derived within
the framework of light-front QCD and the form (the spinor structure
) of the effective interactions is fixed by the lowest order of
light-front QCD. 2) The mass eigen equation is for the squared rest
mass,  the separation between kinematical energy operator and
interaction operators is rigorous. 3)The spinor structure of the
effective interaction make it momentum-energy dependent. 4) Also due
to the spinor structure of the effective interactions, the dynamics
of spin-spin, spin-orbital, and tensor interactions ( especially the
spin singlet-triplet mxing and orbital angular momentum mixing ) are
included( see Appendix C,D ). 5) The predictive power and the
descriptive precision of the model are much better.

 3. Comparing to holographic light-font QCD model of Brodsky $et\ al.$\cite{Brod},
our model has the following new aspects : 1) In the effective
Hamiltonian of mesons, the kinematical energy operator is identical
for both holographic light-front QCD model and our model, but the
interaction terms are quite different. 2) Holographic light-front
QCD model does not specify the effective interaction in detail, but
just simulates confining potential by boundary condition ( or
harmonic oscillator potential), or recently by a positive-sign
dilaton metric to generate confinement and break conformal symmetry;
instead, our model provides a detailed semi-phenomenological
effective interaction including its spinor structure, the confining
potential, and the flavor mixing interactions. 3) Holographic
light-front QCD model does not include spin-spin, spin-orbital, and
tensor interactions, the total angular momentum of the system is not
treated properly ( although it has potential to describe the spin
splittings ); in the contrary, our model specifies the spin
interactions and the spin dynamics is described fully in total
angular momentum representation. 4) Finally, our model has been
applied to a larger number of mesons ( 265 mesons identified
experimentally ) with higher precision than those of holographic
light-front QCD model. In the above respects, our model has provided
a tentative and effective solution to the problems listed above and
the results are amazingly in good agreement with experimental data.
In this sense, our model can be considered to be of complementarity
to and refinement of the holographic light-front QCD model.

This work was supported in part by the National Natural Science
Foundation of China under grant Nos.10974137 and 10775100, and by
the Fund of Theoretical Nuclear Physics Center of HIRFL of China.

\appendix

\section{Dynamics in light front form and instant form in center of mass frame and in internal Hilbert subspace}
To avoid misunderstanding of light-front dynamics, we start from a
discussion of full contents of dynamics for both instant form(IF)
and light front form(LF). The content of dynamics should contain the
following four aspects, we list them for both dynamics of instant
form and dynamics of light form as follows.

\subsection{Full contents of dynamics in instant form}
1) Definition of time $x^0=ct=t (c=1)$ \\

2) Hamiltonian (energy) operator is defined as the time translation
operator:
\begin{eqnarray}
i\hbar\frac{\partial}{\partial x^0}\sim \hat{P}^0=\hat{H}=\hat{M}
\end{eqnarray}
$\hat{M}$ is dynamical mass operator.

3) Dynamics

 (i) Time evolution dynamics: equation of motion
(Schr$\ddot{o}$dinger equation),
\begin{eqnarray}
i\hbar\frac{\partial \Psi}{\partial x^0}=\hat{H}\Psi=\hat{M}\Psi
\end{eqnarray}

 (ii) Stationary dynamics: for stationary solution ,
\begin{eqnarray}
\Psi(t)=e^{-iMt/\hbar}\Psi
\end{eqnarray}
one has the Hamiltonian eigen equation
\begin{eqnarray}
\hat{H}\Psi=\hat{M}\Psi=M\Psi,
\end{eqnarray}
where $M$ is the eigen value of $\hat{M}$.

4) Specification of dynamical operators and kinematical operators
   among Poincare generators:
  6 kinematical operators: $\hat{P}^i,\hat{J}^i,(i=1,2,3)$;
  4 dynamical operators:  $\hat{P}^0, \hat{K}^i,(i=1,2,3)$.

  It should be noted that the dynamical operators contain interactions via
the Hamiltonian and Lorentz boost operators while the kinematical
operators do not. Consequently, the kinematical operators can be
used to characterize the state  of the system as good quantum
numbers according their algebraic structure and the dynamical
operators except the Hamiltonian operator can not play such a role.
It should be emphasized that the above specification is made in
whole Hilbert space of the states of composite systems. For a
composite many-body system, the whole Hilbert space of states can be
factorized into two parts: a) the center of mass motion
characterized by its momentum $\vec{P}$ , and b) the internal motion
characterized by internal quantum numbers and ($J,J^3$ ).
Correspondingly, the Poincare operators contain two kinds of
operations, one on the subspace of center of mass motion and the
other on the subspace of internal motion. Since the center of mass
motion can always be separated from the internal motion, the state
wave function of the composite system $\Psi$  can be written as
$\Psi=\Psi_{cm}\Psi_{inter}$, where the wave function of center of
mass motion is characterized by the center of mass momentum, namely
$\Psi_{cm}=\Psi_{\vec{P}}$ with
$\hat{P}^i\Psi_{\vec{P}}=P^i\Psi_{\vec{P}}$ , while the internal
wave function is characterized by internal quantum numbers and
($J,J^3$ ).

\subsection{Full contents of dynamics in light front form:}
1) Definition of time $x^+$: $x^+=x^0+x^3$

2) Hamiltonian ("energy") operator is defined as the time
translation operator:
\begin{eqnarray}
i\hbar\frac{\partial}{\partial x^+}\sim \hat{P}^-
\end{eqnarray}
From
\begin{eqnarray}
\hat{P}^+\hat{P}^--\hat{P}^2_\bot=\hat{P}_0\hat{P}^0-\hat{P}_3^2-
\hat{P}_1^2-\hat{P}_2^2=\hat{M}_0^2
\end{eqnarray}
$\hat{M}_0$ is rest mass operator; one has
\begin{eqnarray}
\hat{P}^-=\frac{1}{\hat{P}^+}(\hat{M}_0^2+\hat{P}_\bot^2)
\end{eqnarray}

3) Dynamics

(i) Time evolution dynamics: equation of motion
(Schr$\ddot{o}$dinger equation),
\begin{eqnarray}
i\hbar\frac{\partial}{\partial
x^+}\Psi=\frac{1}{\hat{P}^+}(\hat{M}_0^2+\hat{P}_\bot^2)\Psi
\end{eqnarray}

(ii) Stationary dynamics: for stationary solution
$\Psi(\frac{M^2}{P^+},P^+,\vec{P}_\bot,x^{+})$ with quantum numbers:
"energy" $E^{-}=\frac{M^2}{P^+}$ and momentum
$\vec{P}=(P^+,\vec{P}_\bot)$($E^-$ is the eigen value of
$\hat{P}^-$, $P^+$ and $\vec{P}_\bot$ are eigen values of
$\hat{\vec{P}}^+$, $\hat{\vec{P}}_\bot$),
\begin{eqnarray}
\Psi(x^+)=e^{-iM^2x^+/P^+\hbar}\Psi(\frac{M^2}{P^+},P^+,\vec{P}_\bot)
\end{eqnarray}
One has mass eigen equation:
\begin{eqnarray}
\frac{M^2}{P^+}\Psi(\frac{M^2}{P^+},P^+,\vec{P}_\bot)=
\frac{1}{P^+}(\hat{M}_0^2+\hat{P}_\bot^2
)\Psi(\frac{M^2}{P^+},P^+,\vec{P}_\bot)
\end{eqnarray}
or
\begin{eqnarray}
\hat{M}_0^2\Psi(\frac{M^2}{P^+},P^+,\vec{P}_\bot)=
(M^2-\vec{P}_\bot^2)\Psi(\frac{M^2}{P^+},P^+,\vec{P}_\bot)
\end{eqnarray}

4) Specification of dynamical operators and kinematical operators
   among Poincare generators:
  7 kinematical operators: $\hat{P}^+$, $\hat{J}^3$, $\hat{P}^i(i=1,2)$,$\hat{K}^3 $,
  $\hat{E}^1=\hat{K}^1+\hat{J}^2$, $\hat{E}^2=\hat{K}^2-\hat{J}^1 $;
  3 dynamical operators: $\hat{P}^-$, $\hat{F}^1=\hat{K}^1-\hat{J}^2$, $\hat{F}^2=\hat{K}^2+\hat{J}^1 $.

\subsection{Dynamics in center of mass frame and in internal Hilbert subspace for both forms of dynamics }

The internal structure of a composite system should be described in
the rest frame as well as  in the corresponding internal Hilbert
subspace. Since the center of mass frame always follows the center
of mass motion of the system and the position of the center of mass
of the system is at the origin of the frame,  the wave function of
center of mass motion of the system should be $\Psi_{\vec{P}=0}$,
and the center of mass momentum  and the center of mass coordinates
of the system should be zero, namely
$<\Psi_{\vec{P}=0}|\hat{P}^i|\Psi_{\vec{P}=0}>=0$ and
$<\Psi_{\vec{P}=0}|\hat{x}^i|\Psi_{\vec{P}=0}>=0$. In the center of
mass frame, the Hilbert subspace of center of mass motion is frozen
to $\Psi_{\vec{P}=0}$, the whole Hilbert space of states of the
system is thus projected onto the corresponding internal Hilbert
subspace $\Psi_{inter}$. Consequently, the dynamics of the composite
system is reduced to the internal dynamics. Projecting onto the
frozen center of mass wave function and integrating out the center
of mass degrees of freedom, one obtain the Poincare operators in the
internal subspace $\Psi_{inter}$ as follows.

1) Four momentum and property of time in center of mass frame and in
internal Hilbert subspace.

In center of mass frame, the wave function of center of mass motion
is : $\Psi_{\vec{P}=0}$. The four momentum operator in internal
Hilbert subspace can be obtained by projecting out the center of
mass degrees of freedom ( namely averaging over the center of mass
wave function). Since
$\hat{P}^0=\hat{H}=(\hat{M}_{0}^2+\hat{\vec{P}}^2)^{1/2}$ and
$\hat{P}^i\Psi_{\vec{P}=0}=0$, in internal Hilbert subspace, one has

\begin{equation}
\hat{P}_{inter}^i=\left<\Psi_{\vec{P}=0}|\hat{P}^i|\Psi_{\vec{P}=0}\right>=0
\end{equation}

\begin{equation}
\hat{P}_{inter}^0=\left<\Psi_{\vec{P}=0}|\hat{P}^0|\Psi_{\vec{P}=0}\right>=\hat{M}_0.
\end{equation}

Here $\hat{M}_0$ is the operator of rest mass of the system.

 Thus  in internal Hilbert subspace,  the four momentum operators
for instant form read:
\begin{eqnarray}
\hat{P}_{inter}^{\mu}=(\hat{M}_0,0,0,0)
\end{eqnarray}
while  four momentum operators for light front form are:
\begin{eqnarray}
\hat{P}_{inter}^{\mu}=(\hat{M}_0,0,0,\hat{M}_0),
\end{eqnarray}

From the above results , one has
\begin{eqnarray}
\hat{P}_{inter}^-=\hat{P}_{inter}^0=\hat{P}_{inter}^+=\hat{M}_0,
\end{eqnarray}

\begin{eqnarray}
 i\hbar\frac{\partial}{\partial
x^+}=i\hbar\frac{\partial}{\partial
x^0}=i\hbar\frac{\partial}{\partial \tau},
\end{eqnarray}
where $\tau$ is the proper time corresponding to the rest mass
operator $\hat{M}_0$. The last equation leads to
\begin{eqnarray}
x^+=\tau+\tau_0, ~ x^0=\tau+\tau_0'
\end{eqnarray}
where $\tau_0$ and $ \tau_0'$ are constant shifts of proper time.
One can choose the start point of time such that
\begin{eqnarray}
\tau=0 \rightarrow x^+=x^0=0
\end{eqnarray}
This leads to
\begin{eqnarray}
\tau_0=\tau_0'=0
\end{eqnarray}
and
\begin{eqnarray}
x^+=x^0=\tau
\end{eqnarray}

2) Dynamics in center of mass frame  and in internal Hilbert
subspace

 (i) Time evolution dynamics:  equations of motion in center of
mass frame and in internal Hilbert subspace

The Schr$\ddot{o}$dinger equations
\begin{eqnarray}
i\hbar\frac{\partial \Psi}{\partial x^+}=\hat{M}_0\Psi
\end{eqnarray}
in light front form,  and
\begin{eqnarray}
i\hbar\frac{\partial \Psi}{\partial x^0}\Psi=\hat{M}_0\Psi
\end{eqnarray}
in instant form become the same
\begin{eqnarray}
i\hbar\frac{\partial \Psi}{\partial \tau}=\hat{M}_0\Psi
\end{eqnarray}

(ii) Stationary dynamics:  mass (energy) eigen equations in center
of mass frame and in internal Hilbert subspace.

The mass eigen equations
\begin{eqnarray}
\hat{M}_0\Psi=M_0\Psi
\end{eqnarray}
in instant form where $M_0$ is the eigen value of $\hat{M}_0$, and
\begin{eqnarray}
\hat{M}_0^2\Psi=M_0^2\Psi
\end{eqnarray}
in light front form are also the same because multiplying
$\hat{M}_0$ on the first equation leads to the second one.

3) Kinematical and dynamical operators in center of mass frame and
in internal Hilbert subspace

  Projecting onto internal Hilbert subspace, the kinematical and
  dynamical operators can be obtained from the following
  calculation. From the results of (A12-A15) and
\begin{eqnarray}
  \hat{J}^i=\hat{J}_{cm}^i+\hat{J}_{inter}^i,\\
 \hat{J}_{cm}^1=\hat{x}^2\hat{P}^3-\hat{P}^2\hat{x}^3, cyclic,
\end{eqnarray}
one obtain
\begin{eqnarray}
\left<\Psi_{\vec{P}=0}|\hat{J}_{cm}^i|\Psi_{\vec{P}=0}\right>=0,
\end{eqnarray}
\begin{eqnarray}
\left<\Psi_{\vec{P}=0}|\hat{J}^i|\Psi_{\vec{P}=0}\right>=\hat{J}_{inter}^i
\end{eqnarray}

\begin{eqnarray}
&&\left<\Psi_{\vec{P}=0}|\hat{K}^i|\Psi_{\vec{P}=0}\right>
=\left<\Psi_{\vec{P}=0}|\hat{x}^0\hat{P}^i-\hat{x}^i\hat{P}^0|\Psi_{\vec{P}=0}\right>,\\
\nonumber
&&=\left<\Psi_{\vec{P}=0}|\hat{x}^i|\Psi_{\vec{P}=0}\right>\hat{M}_0=0,
\end{eqnarray}

\begin{eqnarray}
\left<\Psi_{\vec{P}=0}|\hat{E}^1|\Psi_{\vec{P}=0}\right>=
\left<\Psi_{\vec{P}=0}|\hat{K}^1+\hat{J}^2|\Psi_{\vec{P}=0}\right>=\hat{J}_{inter}^2
\end{eqnarray}

\begin{eqnarray}
\left<\Psi_{\vec{P}=0}|\hat{E}^2|\Psi_{\vec{P}=0}\right>=
\left<\Psi_{\vec{P}=0}|\hat{K}^2-\hat{J}^1|\Psi_{\vec{P}=0}\right>=-\hat{J}_{inter}^1
\end{eqnarray}

\begin{eqnarray}
\left<\Psi_{\vec{P}=0}|\hat{F}^1|\Psi_{\vec{P}=0}\right>=
\left<\Psi_{\vec{P}=0}|\hat{K}^1-\hat{J}^2|\Psi_{\vec{P}=0}\right>=-\hat{J}_{inter}^2
\end{eqnarray}

\begin{eqnarray}
\left<\Psi_{\vec{P}=0}|\hat{F}^2|\Psi_{\vec{P}=0}\right>=
\left<\Psi_{\vec{P}=0}|\hat{K}^2+\hat{J}^1\hat{P}^0|\Psi_{\vec{P}=0}\right>=\hat{J}_{inter}^1
\end{eqnarray}

From the above results, one obtains the same reduced and degenerated
kinematical and dynamical operators  for both forms of dynamics in
internal Hilbert subspace as follows: kinematical operators:
$\hat{J}_{inter}^i(i=1,2,3)$; dynamical operator:
$\hat{P}_{inter}^0=\hat{P}_{inter}^-=\hat{M}_0$.

The above results tell that in center of mass frame and in internal
Hilbert subspace, light front time and  instant time, light front
dynamics and instant dynamics, light front angular momentum and
instant angular momentum are identical.

\subsection{Conclusion}
 In general frames and in whole Hilbert space,  both forms of dynamics
are quite different. However, in center of mass frame and in
internal Hilbert subspace, the  two forms of dynamics are reduced to
the identical internal dynamics.

There is a dilemma in this paper at first glance: our model begins
with a light front QCD model, but the final form of our model
possesses the feature of instant dynamics of QCD. Is it of LF
dynamics or IF dynamics? The solution to the dilemma is given in
this Appendix, the answer is that in center of mass frame and in
internal Hilbert subspace, the reduced internal dynamics of both
forms are identical.

Therefore, our model contains ingredients of both the instant form
and light front form of QCD, it can be called as QCD inspired
effective Hamiltonian meson model.

\section{Conservation of total angular momentum in  internal Hilbert subspace}

The reduction of angular momentum operators in internal Hilbert
subspace can be discussed in an alternative manner and the results
are the same as that in Appendix A.

A relativistic dynamical system has inhomogeneous Lorentz symmetry
defined by the $Poincar\acute{e}\ algebra$: $P^\mu$ is
energy-momentum vector, and $M^{\mu\nu}$ is used to describes the
rotational and boost transformations. In instant form, the angular
momentum and boost vectors are given as: $M^{ij}=\epsilon_{ijk}J^k$
and $M^{0i}=K^i$

Now define the "quasi angular momentum" operators in the light-front
form:
\begin{eqnarray}
& &\mathcal {J}^3 =J^3+\frac{\varepsilon_{ij}{\bm E}_\perp^i {\bm P}
_\perp^j}{P^+}  \ , \nonumber  \\
& &\mathcal {J}^{\perp
i}=M_{0}^{-1}\varepsilon_{ij}(\frac{1}{2}({\bm F}_\perp^j P^+ - {\bm
E}_\perp^j P^-) - K^3 {\bm P}_\perp^j
\nonumber \\
& & +\mathcal {J}^3\varepsilon_{jl}{\bm P}_\perp^l),
 (i,j=1,2).
\end{eqnarray}

It is easy to prove that they satisfy the SU(2) algebra:
\begin{eqnarray}
[\mathcal {J}^i,\mathcal {J}^j]=i \epsilon_{ijk}\mathcal {J}^k
\end{eqnarray}

It is very useful to define a `light-front Hamiltonian' as the
operator:
\begin{eqnarray}
H_{LC}=P^\mu P_\mu=P^-P^+-\vec{P_\bot}^2=\hat{M}_0^2
\end{eqnarray}

$H_{LC}$  commutes with the quasi angular momentum operators :

\begin{eqnarray}
[H_{LC},\vec{\mathcal {J}}]=0 .
\end{eqnarray}
In principle, one could label the eigen states as $|M,P^+,
\vec{P_\bot}, \vec{\mathcal{J}}^2, \vec{\mathcal {J}}_3 \rangle$,
since $\mathcal {J}_3$ is kinematical. However, $\vec{\mathcal
{J}_\bot}$ is dynamical and depends on the interactions. Thus it is
generally difficult to explicitly compute the total spin
$\vec{\mathcal{J}}$ of a state using light-front quantization.
Fortunately, in center-of-mass frame and in internal Hilbert
subspace, by using the results of Appendix A, one has the following
equations,
\begin{eqnarray}
& &\left<\Psi_{\vec{P}=0}|\mathcal {J}^3|\Psi_{\vec{P}=0}\right> =J_{inter}^3  \ , \nonumber  \\
& &\left<\Psi_{\vec{P}=0}|\mathcal {J}^{i}
|\Psi_{\vec{P}=0}\right>=J_{inter}^i\\
& &\qquad\qquad\qquad\qquad\qquad\qquad  (i,j=1,2) \ \, . \nonumber
\end{eqnarray}

Therefore, in internal Hilbert subspace, the quasi angular momentum
operators $\mathcal {J}_{inter}^i$  are identical to the total
angular momentum operators $J_{inter}^i(i=1,2,3)$, the total angular
momentum is conserved, and the eigen equation of the Hamiltonian
$H_{LC}$ of the internal dynamics can be solved in the total angular
momentum representation.

\section{Derivation of the radial mass eigen equations in total angular momentum representation}

According to Pauli et al., the effective mass eigen equation of
mesons of light-front QCD in center of mass frame and in internal
Hilbert subspace reads:

\begin{eqnarray}
\label{eq:pauli01}
 \lefteqn{ \hspace*{-1cm}
  \left[ M_0^{2} - \left( E_{1}(k)+E_{2}(k) \right)^{2}
  \right] \varphi_{s_{1}s_{2}}(\bm{k})
 }
 \nonumber \\
 & = &
 \sum_{s_{1}^{\prime}s_{2}^{\prime}} \int d^{3}\bm{k}
 U_{s_{1}s_{2};s_{1}^{\prime}s_{2}^{\prime}}(\bm{k};\bm{k^\prime})
 \varphi_{s_{1}^{\prime }s_{2}^{\prime}}(\bm{k}^{\prime}),
 \label{eq:main}
\end{eqnarray}
where
\begin{equation}
 U_{s_{1}s_{2};s_{1}^{\prime}s_{2}^{\prime}}
 =
 \displaystyle\frac{4m_{s}}{3\pi ^{2}}
 \displaystyle\frac{\overline{\alpha}(Q)}{Q^{2}} R(Q)
 \displaystyle\frac{S_{s_{1}s_{2};s_{1}^{\prime}s_{2}^{\prime}}}
                   {\sqrt{A(k)A(k^{\prime})}}
\end{equation}
with
\begin{eqnarray}
 S_{s_{1}s_{2};s_{1}^{\prime}s_{2}^{\prime}}
 & = &
 [ \overline{u}( \bm{k},s_{1}) \gamma_{\mu}(1) u( \bm{k^\prime},s_{1}^{\prime}) ]
 \nonumber \\
 & \times &
 [ \overline{v}(-\bm{k},s_{2}) \gamma^{\mu}(2) v(-\bm{k^\prime},s_{2}^{\prime}) ]
\end{eqnarray}
and
\begin{eqnarray}
 \displaystyle\frac{1}{A(k)}
 & = &
 m_{r} \left( \displaystyle\frac{1}{E_{1}(k)}
             +\displaystyle\frac{1}{E_{2}(k)}
       \right) ,\ \nonumber
 \\
 m_{s}
 & = &
 m_{1}+m_{2},\ m_{r} = \displaystyle\frac{m_{1}m_{2}}{m_{1}+m_{2}},\
 \nonumber\\
 Q
 & = &
 Q(\bm{k};\bm{k}^{\prime}).
\end{eqnarray}

Equation~(A1) can be written as Schr\"{o}dinger equation in the
light front QCD,
\begin{equation}
\label{eq:B5}
 \widehat{H} \Psi_{\text{meson}} = M_0^{2} \Psi_{\text{meson}}
\end{equation}
The general eigen wave function $\Psi_{\text{meson}}$ of meson can
be expressed in momentum-spin representation,
\begin{equation}
\label{eq:B6}
 \Psi_{\text{meson}}
 = \sum_{s_{1},s_{2}} \int d^{3}\bm{k} \varphi_{s_{1}s_{2}}(\bm{k})
   \left\vert \chi(s_{1})\chi(s_{2}) \cdot \bm{k} \right \rangle
 .
\end{equation}

Here basis of the momentum-spin representation  are
\begin{equation}
 \langle \bm{r} | \chi(s_{1})\chi(s_{2}) \cdot \bm{k} \rangle
 = \displaystyle \frac{1}{(2\pi\hbar)^{3/2}}
   \chi(s_{1})\chi(s_{2})e^{i\bm{k}\cdot \bm{r}},
\end{equation}
where the spin wave functions and their orthogonal conditions read
\begin{equation}
 \chi(+\frac{1}{2}) = \left(
                       \begin{array}{c}
                        1 \\
                        0
                       \end{array}
                      \right) ,\
 \chi(-\frac{1}{2}) = \left(
                       \begin{array}{c}
                        0 \\
                        1
                       \end{array}
                      \right) ,
\end{equation}
\begin{equation}
 \langle \chi(s_{1}) | \chi(s_{2}) \rangle = \delta_{s_{1}s_{2}}.
\end{equation}
The orthogonal conditions of the spinors are
\begin{eqnarray}
 \lefteqn{ \hspace*{-3.0cm}
  \langle
   \bm{k} \cdot \overline{u}(\bm{k},s_{1}) \overline{v}(-\bm{k},s_{2}) |
   u(\bm{k}^{\prime},s_{1}^{\prime}) v(-\bm{k}^{\prime},s_{2}^{\prime})
   \cdot \bm{k}^{\prime}
  \rangle
 }
 \nonumber \\
 & = &
 \delta^{(3)}(\bm{k}-\bm{k}^{\prime})
 \delta_{s_{1}s_{1}^{\prime}}
 \delta_{s_{2}s_{2}^{\prime}},
 \\
 \hspace*{-0.6cm}
 \overline{u}(\bm{k},s_{1}) u(\bm{k},s_{1}^{\prime})
 & = &
 \delta_{s_{1}s_{1}^{\prime}},
 \\
 \hspace*{-0.6cm}
 \overline{v}(-\bm{k},s_{2}) v(-\bm{k},s_{2}^{\prime })
 & = &
 \delta_{s_{2}s_{2}^{\prime}},
\end{eqnarray}
and the completeness conditions read,
\begin{eqnarray}
 \sum_{s} u( \bm{k},s) \overline{u}( \bm{k},s)
 & = &\frac{1}{2m}
 \left( \gamma_{\mu} k_{1}^{\mu} +m \right)
 , \\
 \sum_{s} v(-\bm{k},s) \overline{v}(-\bm{k},s)
 & = &\frac{1}{2m}
 \left( \gamma_{\mu} k_{2}^{\mu} -m \right) ,
\end{eqnarray}
where $k_{1}^{\mu}=(E_{1}(k),\bm{k})$, and
$k_{2}^{\mu}=(E_{2}(k),-\bm{k})$.

According to the Dirac form of quantum mechanics, in the eigen
equation (\ref{eq:B5}), the Dirac form of the Hamiltonian operator
is
\begin{equation}
\label{eq:B15}
 \widehat{H} = \widehat{E} + \widehat{U},
\end{equation}
where
\begin{eqnarray}
\label{eq:B16}
 \widehat{E} & = &
 \int d^{3} \bm{k}
 \displaystyle \left[ E_{1}(k)+E_{2}(k) \right]^{2}
 \nonumber \\
 &   & \mbox{} \times
 \sum_{s_{1}s_{2}} |\chi(s_{1}) \chi (s_{2}) \cdot \bm{k} \rangle
 \langle \bm{k} \cdot \chi(s_{1}) \chi(s_{2})| ,
\end{eqnarray}
and
\begin{eqnarray}
\label{eq:B17}
 \lefteqn{
  \widehat{U} =
  \int d^{3}\bm{k} d^{3}\bm{k}^{\prime}
  \sum_{s_{1}s_{2};s_{1}^{\prime}s_{2}^{\prime}}
  U(k,k^{\prime})
 }
 \nonumber \\
 &   & \mbox{} \times
 \left[ \overline{u}(\bm{k},s_{1}) \overline{v}(-\bm{k},s_{2})
        (\gamma_{\mu}(1) \gamma^{\mu}(2))
        u(\bm{k}^{\prime},s_{1}^{\prime}) v(-\bm{k}^{\prime},s_{2}^{\prime})
 \right]
 \nonumber \\
 &   & \mbox{}
 \times | \chi(s_{1}) \chi(s_{2}) \cdot \bm{k} \rangle
        \langle \bm{k}^{\prime} \cdot \chi(s_{1}^{\prime})\chi(s_{2}^{\prime})
        |,
\end{eqnarray}
with the definition,
\begin{equation}
\label{eq:B18}
 U(k,k^{\prime})
 \equiv
 \displaystyle\frac{1}{3m_{r}\pi^{2}}
 \displaystyle\frac{\overline{\alpha}(Q)}{Q^{2}}R(Q)
 \displaystyle\frac{1}{\sqrt{A(k)A(k^{\prime})}}.
\end{equation}

In the above equation, as done by Pauli et al.\cite{Pauli03}, the
light front $k-$ space has been transformed back to the Lab $k-$
space by the Terent'ev transformation, and Lepage-Brodsky (helicity)
spinors have been transformed to the Bjorken-Drell (spin) spinors.

Using eqs.(\ref{eq:B6}, \ref{eq:B15}-\ref{eq:B18}) and projecting
equation (\ref{eq:B5}) onto the subspace $ |\chi(s_{1}) \chi(s_{2})
\cdot \bm{k} \rangle$, we recover the equation (\ref{eq:pauli01}),
indicating that the Dirac Form of the eigen equation (\ref{eq:B5})
is equivalent that of (\ref{eq:pauli01}).

  Since $(E_{1}(k)+E_{2}(k))^{2}$ and the interaction kernal operator
  $\widehat{U}[\bm k,\bm k^{\prime};\bm\sigma(1),\bm\sigma(2)] $
 are scalar (see Appendix D, discussion below eq.(\ref{eq:85})),
$\widehat{H}$ is rotational invariant with respect to the total
angular momentum $\bm J_{i} = \bm l_{i} + \bm s_{i}^{1} + \bm
s_{i}^{2} = \bm l_{i} + \bm s_{i}$, $[\widehat{H},\bm J_{i}] = 0$.
That means the total angular momentum $\widehat{\bm J}^{2}$ and
$\widehat{J}_{z}$ are conserved. Based on this point, the wave
function of the meson system can be written in total angular
representation as follows,
\begin{equation}
\label{eq:B19}
 \Psi_{meson}(k,\Omega_{k},s) = \sum_{J,M}\sum_{l=|J-s|}^{J+s} \sum_{s=0,1}R_{Jsl}(k)
 \Phi_{JslM}(\Omega_{k},s),
\end{equation}
were the total angular momentum eigen functions $\Phi_{JslM}$ of $\{
\widehat{\bm J}^{2}, \widehat{J}_{z},\widehat{\bm
s}^{2},\widehat{\bm l}^{2} \}$ are,
\begin{equation}
 \Phi_{JslM}(\Omega_{k},s)
 = \sum_{m\mu} \langle lms\mu |JM \rangle
   Y_{lm}(\Omega_{k}) \chi_{s\mu}(12),
\end{equation}
the eigen wave functions of spin singlet and triplet read as,
\begin{equation}
 \chi_{s\mu}(12) = \sum_{s_{1}s_{2}}
 \textstyle \langle\frac{1}{2}s_{1}\frac{1}{2}s_{2} | s\mu \rangle
 \chi(s_{1}) \chi(s_{2}) .
\end{equation}

By virtue of the Fourier transformation in spherical coordinates,
from the eigen wave function in the momentum radial $k-$space, one
can obtain the corresponding wave function in  the configuration
radial $r-$ space,
\begin{eqnarray}
 \Psi_{JM}(r,\Omega_{r},s)
 & = & \int  d\bm{k}^{3}\Psi_{JM}(k,\Omega_{k},s) e^{i\bm{k}\cdot\bm{r} }
 \nonumber \\
 & = & \sum_{l=|J-s|}^{J+s} \sum_{s=0,1}\sum_{l',m'} \int k^{2}dk
 R_{Jsl}(k)J_{l}(kr)
 \nonumber \\
 & &\int d\Omega_{k} \Phi_{JslM}(\Omega_{k},s)Y^{*}_{l'm'}(\Omega_{k})Y_{l'm'}(\Omega_{r})
 \nonumber \\
 & = & \sum_{l=|J-s|}^{J+s}\sum_{s=0,1}R_{Jsl}(r)\Phi_{JslM}(\Omega_{r},s),
\end{eqnarray}
 where
\begin{eqnarray}
 \Phi_{JslM}(\Omega_{r},s)
 & = &
 \sum_{m\mu} \langle lms\mu |JM \rangle
  Y_{lm}(\Omega_{r})\chi_{s\mu}(12),
 \nonumber \\
 R_{Jsl}(r) & = & \int k^{2} dk R_{Jsl}(k) J_{l}(kr) ,
 \nonumber \\
 J(kr) & = & \sqrt{4\pi (2l+1)}\ i^{l}\ j_{l}(kr) .
\end{eqnarray}
$j_{l}(kr)$ is the spherical Bessel function of order $l$.

Using the expression (\ref{eq:B19}) of the wave function
$\Psi_{meson}$, projecting the mass eigen equation (\ref{eq:B5})
onto the $\Phi_{JslM}$ subspace from the left, and integrating out
the spin and angular part of the wave function, we obtain the eigen
equations for the radial wave functions $R_{Jsl}(k)$,
\begin{eqnarray}
 \lefteqn{
  \left[ M_0^{2}-\left( E_{1}(k)+E_{2}(k)\right) ^{2}\right] R_{Jsl}(k)
 }
 \\
 & = &
  \sum_{l^{\prime}=|J-s'|}^{J+s'} \sum_{s'=0,1} \int k^{\prime 2}dk^{\prime}
   U_{sl;s^{\prime} l^{\prime}}^{J}(k;k^{\prime})
   R_{Js'l^{\prime}}(k^{\prime}) ,
 \nonumber
\end{eqnarray}
where the kernel $U_{Sl;S^{\prime}l^{\prime}}^{J}(k;k^{\prime})$ is
defined as,
\begin{eqnarray}
 \lefteqn{
  U_{sl;s^{\prime}l^{\prime}}^{J}(k;k^{\prime})
  =
  \sum_{mm^{\prime}} \sum_{s_{1}s_{2}} \sum_{s_{1}^{\prime}s_{2}^{\prime}}
  \int \int d\Omega_{k} d\Omega_{k^{\prime}}
 }
 \nonumber \\
 &   & \hspace*{-0.2cm} \mbox{}\times
 \langle Y_{lm}(\Omega_{k}) |
         U_{s_{1}s_{2};s_{1}^{\prime}s_{2}^{\prime}}(\textbf{k},\textbf{k}^{\prime})
         | Y_{l^{\prime}m^{\prime}}(\Omega_{k^{\prime}})
 \rangle
 \\
 &   & \hspace*{-0.2cm} \mbox{} \times
 \langle lms\mu |JM \rangle
 \textstyle\langle \frac{1}{2}s_{1}\frac{1}{2}s_{2} | s\mu\rangle
 \langle l^{\prime}m^{\prime}s^{\prime}\mu^{\prime} | JM\rangle
 \textstyle\langle \frac{1}{2}s_{1}^{\prime}\frac{1}{2}s_{2}^{\prime} |
                   s^{\prime}\mu^{\prime}
           \rangle .
 \nonumber
\end{eqnarray}
This is a set of coupled equations for the radial functions
$R_{Jsl}(k)$ that have different partial waves, spin singlet and
triplet coupled by the tensor potentials and by the relativistic
spin-orbital potential (see below).

\section{Calculation of the interaction kernel in total angular momentum representation}
The quark and anti-quark spinors are given in the Bj\o rken-Drell
representation,
\begin{eqnarray*}
 u(\bm{k},s=+\textstyle\frac{1}{2})
 & = &
  \displaystyle \frac{1}{\sqrt{2m_{1}(E_{1}+m_{1})}}
   \left(
    \begin{array}{c}
     E_{1}+m_{1} \\
     0 \\
     k_{z} \\
     k_{l}
    \end{array}
   \right) ,
 \nonumber \\
 u(\bm{k},s=-\textstyle\frac{1}{2})
 & = &
  \displaystyle\frac{1}{\sqrt{2m_{1}(E_{1}+m_{1})}}
   \left(
    \begin{array}{c}
     0 \\
     E_{1}+m_{1} \\
     k_{r} \\
    -k_{z}
   \end{array}
  \right) ,
 \nonumber \\
 v(-\bm{k},s=+\textstyle\frac{1}{2})
 & = &
  \displaystyle\frac{1}{\sqrt{2m_{2}(E_{2}+m_{2})}}
   \left(
    \begin{array}{c}
    -k_{z} \\
    -k_{l} \\
     E_{2}+m_{2} \\
     0
    \end{array}
   \right) ,
 \nonumber \\
 v(-\bm{k},s=-\textstyle\frac{1}{2})
 & = &
  \displaystyle\frac{1}{\sqrt{2m_{2}(E_{2}+m_{2})}}
   \left(
    \begin{array}{c}
    -k_{r} \\
     k_{z} \\
     0 \\
     E_{2}+m_{2} \\
    \end{array}
   \right) ,
 \nonumber
\end{eqnarray*}
where
\begin{eqnarray}
 k_{l,r} & = & k_{x}\pm ik_{y} =k \sin\theta_{k} e^{\pm i\varphi_{k}}
         = k \sqrt{\frac{8\pi}{3}} Y_{1\pm 1}(\theta_{k},\varphi_{k}),
 \nonumber \\
 k_{z} & = & k \cos\theta_{k}
         = k \sqrt{\frac{4\pi}{3}} Y_{10}(\theta_{k},\varphi_{k}).
\end{eqnarray}
Defining the spherical spinors
\begin{equation}
 \Phi_{\frac{1}{2}s}^{A}(\Omega_{k})
  = \sum_{m\nu}
     \textstyle\langle 1m \frac{1}{2}\nu | \frac{1}{2}s \rangle
             Y_{00}(\Omega_{k}) \chi(\nu)
  = \displaystyle\frac{1}{\sqrt{4\pi}} \chi(s),
\end{equation}
\begin{equation}
 \Phi_{\frac{1}{2}s}^{B}(\Omega_{k})
 = \sum_{m\nu}
    \textstyle\langle 1m \frac{1}{2}\nu | \frac{1}{2}s \rangle
             Y_{1m}(\Omega_{k}) \chi(\nu)
  = \displaystyle\frac{1}{\sqrt{4\pi}} \sigma_{k}\chi(s),
\end{equation}
where $\sigma_{k} = (\bm{\sigma} \cdot \bm{k})/k $ and $\Omega_{k} =
(\theta_{k},\varphi_{k})$($\sigma_{k}$ is pseudo scalar ), the
spinors can be re-expressed as
\begin{eqnarray}
 u(\bm{k},s)
 & = &
 \left(
  \begin{array}{c}
   \phantom{-} A_{1}(k)\Phi_{\frac{1}{2}s}^{A}(\Omega_{k}) \\
   \phantom{-} B_{1}(k)\Phi_{\frac{1}{2}s}^{B}(\Omega_{k})
  \end{array}
 \right) \ ,
 \\
 v(-\bm{k},s)
 & = &
 \left(
  \begin{array}{c}
             - B_{2}(k)\Phi_{\frac{1}{2}s}^{B}(\Omega_{k}) \\
   \phantom{-} A_{2}(k)\Phi_{\frac{1}{2}s}^{A}(\Omega_{k})
  \end{array}
 \right) \ ,
\end{eqnarray}
where
\begin{equation}
 A_{i}(k) = \sqrt{\frac{2\pi (E_{i}+m_{i})}{m_{i}}},
 \
 B_{i}(k) = \sqrt{\frac{2\pi k^{2}}{m_{i}(E_{i}+m_{i})}} \
 .
\end{equation}

The spin factor $S_{s_{1}s_{2};s_{1}^{\prime }s_{2}^{\prime }}$ of
the interaction can be written as
\begin{widetext}
\begin{eqnarray}
 S_{s_{1}s_{2};s_{1}^{\prime}s_{2}^{\prime}}
 & = &
 \left[
  \overline{u}( \bm{k},s_{1}) \gamma_{0}(1) u( \bm{k^\prime},s_{1}^{\prime})
 \right]
 \left[
  \overline{v}(-\bm{k},s_{2}) \gamma_{0}(2) v(-\bm{k^\prime},s_{2}^{\prime})
 \right]
 -
 \left[
  \overline{u}( \bm{k},s_{1}) \gamma_{i}(1) u( \bm{k^\prime},s_{1}^{\prime})
 \right]
 \left[
  \overline{v}(-\bm{k},s_{2}) \gamma_{i}(2) v(-\bm{k^\prime},s_{2}^{\prime})
 \right]
 \nonumber \\
 & = &
 \left[
  A_{1}^{\ast}(k) A_{1}(k^{\prime})
 +B_{1}^{\ast}(k) B_{1}(k^{\prime})
   \big\langle\Phi_{\frac{1}{2}s_{1}}^{B}(\Omega_{k}) \big|
              \Phi_{\frac{1}{2}s_{1}^{\prime}}^{B}(\Omega_{k^{\prime}})
   \big\rangle
 \right]
 \left[
  A_{2}^{\ast}(k) A_{2}(k^{\prime})
 +B_{2}^{\ast}(k) B_{2}(k^{\prime})
   \big\langle \Phi_{\frac{1}{2}s_{2}}^{B}(\Omega_{k}) \big|
               \Phi_{\frac{1}{2}s_{2}^{\prime}}^{B}(\Omega_{k^{\prime}})
   \big\rangle
 \right]
 \nonumber \\
 &   & \mbox{} +
 \left[
  A_{1}^{\ast}(k) B_{1}(k^{\prime})
   \big\langle \Phi_{\frac{1}{2}s_{1}}^{A}(\Omega_{k}) \sigma_{i} \big|
               \Phi_{\frac{1}{2}s_{1}^{\prime}}^{B}(\Omega_{k^{\prime}})
   \big\rangle
 +B_{1}^{\ast}(k) B_{1}(k^{\prime})
   \big\langle \Phi_{\frac{1}{2}s_{1}}^{B}(\Omega_{k}) \big| \sigma _{i}
           \Phi_{\frac{1}{2}s_{1}^{\prime}}^{A}(\Omega_{k^{\prime}})
   \big\rangle
 \right]
 \nonumber \\
 &   & \mbox{} \times
 \left[
  B_{2}^{\ast}(k) A_{2}(k^{\prime})
   \big\langle \Phi_{\frac{1}{2}s_{1}}^{B}(\Omega_{k}) \big| \sigma _{i}
               \Phi_{\frac{1}{2}s_{1}^{\prime}}^{A}(\Omega_{k^{\prime}})
   \big\rangle
 +A_{2}^{\ast}(k) B_{2}(k^{\prime})
   \big\langle \Phi_{\frac{1}{2}s_{2}}^{A}(\Omega_{k}) \sigma _{i} \big|
               \Phi_{\frac{1}{2}s_{2}^{\prime}}^{B}(\Omega_{k^{\prime}})
   \big\rangle
 \right]
 \nonumber \\
 & = &
 \frac{1}{\sqrt{4\pi}}
 \Big\langle \chi(s_{1})\chi(s_{2}) \Big|
  \Big\{
   \left[
    A_{1}^{\ast}(k) A_{1}(k^{\prime})
   +B_{1}^{\ast}(k) B_{1}(k^{\prime}) \sigma_{k}(1) \sigma_{k^{\prime}}(1)
   \right]
   \left[
    A_{2}^{\ast}(k) A_{2}(k^{\prime})
   +B_{2}^{\ast}(k) B_{2}(k^{\prime}) \sigma_{k}(2) \sigma_{k^{\prime}}(2)
   \right]
 \nonumber \\
 &   & \hspace*{2.8cm} \mbox{}
  +\left[
    A_{1}^{\ast}(k) B_{1}(k^{\prime}) \bm{\sigma}(1) \sigma_{k^{\prime}}(1)
   +B_{1}^{\ast}(k) A_{1}(k^{\prime}) \sigma_{k}(1) \bm{\sigma}(1)
   \right]
 \nonumber \\
 &   & \hspace*{2.8cm} \mbox{}\cdot
   \left[
    B_{2}^{\ast}(k) A_{2}(k^{\prime}) \sigma_{k}(2) \bm{\sigma}(2)
   +A_{2}^{\ast}(k) B_{2}(k^{\prime}) \bm{\sigma}(2) \sigma_{k^{\prime}}(2)
   \right]
  \Big\} \Big| \chi(s_{1}^{\prime}) \chi(s_{2}^{\prime}) \Big\rangle .
\end{eqnarray}
The kernel $U_{sl;s^{\prime}l^{\prime}}^{J}(k;k^{\prime})$ can be
rewritten as
\begin{equation}
 U_{sl;s^{\prime}l^{\prime}}^{J}(k;k^{\prime})
 = \big\langle \Phi_{JslM}(\Omega_{k},s)
   \big|
    \widehat{U}[\bm k,\bm k^{\prime};\bm\sigma(1),\bm\sigma(2)]
   \big|
    \Phi_{Js^{\prime}l^{\prime}M}(\Omega_{k^{\prime}},s^{\prime})
   \big\rangle,
\end{equation}
where the interaction operator in momentum and spin space is
\begin{eqnarray}
\label{eq:85}
 \widehat{U}[\bm k,\bm k^{\prime};\bm\sigma(1),\bm\sigma(2)]
 =
 \frac{U(k, k^\prime)}{\sqrt{4\pi}} \!\!\!
 & \Big\{ & \!\!
   \left[
    A_{1}^{\ast}(k) A_{1}(k^{\prime})
   +B_{1}^{\ast}(k) B_{1}(k^{\prime}) \sigma_{k}(1) \sigma_{k^{\prime}}(1)
   \right]
   \left[
    A_{2}^{\ast}(k) A_{2}(k^{\prime})
   +B_{2}^{\ast}(k) B_{2}(k^{\prime}) \sigma_{k}(2) \sigma_{k^{\prime}}(2)
   \right]
 \nonumber \\
  &   & \hspace*{2.8cm} \mbox{}
  + \left[
   A_{1}^{\ast}(k) B_{1}(k^{\prime}) \bm{\sigma}(1) \sigma_{k^{\prime}}(1)
  +B_{1}^{\ast}(k) A_{1}(k^{\prime}) \sigma_{k}(1) \bm{\sigma}(1)
  \right]
 \nonumber \\
&   & \hspace*{2.8cm} \mbox{}\cdot
   \left[
   B_{2}^{\ast}(k) A_{2}(k^{\prime}) \sigma_{k}(2) \bm{\sigma}(2)
  +A_{2}^{\ast}(k) B_{2}(k^{\prime}) \bm{\sigma}(2) \sigma_{k^{\prime}}(2)
  \right]
  \Big\}
 .
\end{eqnarray}
\end{widetext}
Since  $\sigma_{k}$ and  $\sigma_{k'}$ are pseudo scalar, $k$, $k'$,
$\sigma_{k}$ $\sigma_{k'}$, and $\bm{\sigma}(1)\cdot \bm{\sigma}(2)$
are scalar, the above interaction kernel operator $\widehat{U}[\bm
k,\bm k^{\prime};\bm\sigma(1),\bm\sigma(2)] $ is scalar.

 From the last expression of the kernel
$U_{sl;s^{\prime}l^{\prime}}^{J}(k;k^{\prime})$, we could see that
the first term contributes to different kinds of central potentials
and relativistic spin-orbit coupling potentials, the second term
contributes to the tensor potentials changing $l$ by $\Delta l =\pm
2$ and mixing spin singlet and triplet.

If $m_{1}=m_{2}$ and the tensor potentials are neglected,  $l$ and
$s$ are conserved and the interaction kernel becomes diagonal in $l$
and $s$ representation,
\begin{equation}
 U_{sl;s^{\prime}l^{\prime}}^{J}(k;k^{\prime})
 = U_{sl;sl}^{J}(k;k^{\prime}) \delta_{ll^{\prime}} \delta_{ss^{\prime}}
 = U_{Jsl}(k;k^{\prime}) \delta_{ll^{\prime}} \delta_{ss^{\prime}} .
\end{equation}

\end{document}